# Within-person prediction of depressive symptom change using year-long Screenome data and CES-D assessments

Merve Cerit, Andrea Mock*, Vryan Almanon Feliciano*, Thomas N. Robinson, Byron Reeves, Nilam Ram, Nick Haber

Stanford University, Stanford, CA, USA

Correspondence: mervecer@stanford.edu

*equal contributions

## Abstract

Predicting whether an individual's depressive symptoms will worsen, remain stable, or improve over the coming weeks can enable earlier and more targeted care, yet prospective within-person trajectory prediction remains largely unaddressed in digital phenotyping. We combine fortnightly CES-D assessments with over 100 million screenshots captured every five seconds via the Stanford Screenomics platform from 96 adults followed for approximately one year (mean 20.9 ± 3.9 assessments per participant, 2,002 total observations). We frame prediction as a within-person classification task: whether symptoms will worsen, remain stable, or improve over the subsequent fortnight, operationalized in three ways to capture clinically meaningful change. Under temporal holdout, XGBoost achieves an AUC of 0.906 for crossings of established CES-D severity bands and 0.755 for change relative to each participant's own within-person variability, generalizing to unseen individuals (AUC = 0.821). Each person's typical symptom level was the only statistically significant predictor above the most recent CES-D score; without it, the most consequential worsening transitions go undetected. Screenome-derived behavioral features revealed prodromal patterns of worsening, including escalating social media use, fragmented device engagement, and changes in overnight activity, with substantial individual heterogeneity. These findings establish a proof-of-concept foundation for monitoring systems that could identify individuals approaching clinical deterioration before symptoms reach a crisis point.

## Introduction

Although depression is not a stable state, screening and care delivery systems often operate as though individuals' experiences of depression are stable. Most people with depression go untreated, and those who do receive care are seen episodically, at intervals that cannot capture the continuous fluctuation of symptoms[1]. The volatility and trajectories of individuals' depressive symptoms differ substantially. Thus, the gaps between clinical encounters are not neutral time simply because they are unobserved; rather, they are periods during which symptoms may worsen, remain stable, or improve[2,3].

Because between-person differences (how individuals differ from each other) do not predict within-person processes (how a given individual's symptoms change over time)[4,5], predicting whether a specific person's symptoms will worsen over the coming weeks requires modeling each person's own symptom history, and not just whether they have more or fewer symptoms than other persons. Anticipating how a person's symptoms will change, rather than confirming where they are at the most recent assessment, can transform when and how care is delivered [6].

Digital phenotyping, using personal digital devices to continuously capture behavioral signals relevant to health, has emerged as a promising approach for both screening individuals who may never reach a clinic and for monitoring symptom trajectories between assessments[7–9]. A decade of work has established that behavioral features passively derived from smartphones (e.g., GPS mobility, phone usage, sleep proxies, and communication logs) correlate with depressive symptom severity, and that within-person designs may outperform those relying solely on between-person differences[10–14]. The Screenomics framework used in this study obtains a rich history of behavioral features by taking screenshots of users' smartphone screens every few seconds, enabling measurement of how much time people spend on their smartphones, how they move between apps, whether they consume or produce content, and the actual visual content displayed on their screen – all at a finer granularity than obtained in digital phenotyping studies of GPS and call-log traces [15–18]. Prior work with screenome data has established strong within-person covariation between mental health assessments and Screenome-derived behavioral features[19], providing the foundation for prospective symptom prediction we address here.

Despite substantial progress, three gaps persist. First, prospective prediction of within-person symptom trajectory direction remains largely unaddressed. Existing longitudinal passive sensing work has focused on predicting between-person differences in symptom severity levels, affect states, or disorder onset [12,13,20,21] rather than predicting whether a specific individual's symptoms will worsen, remain stable, or improve over the coming weeks. Studies spanning one year or longer, or that evaluate temporal generalization within individuals, are scarce[22–25]. Second, there is no consensus on what drives predictive performance in within-person depression models: the relative predictive contribution of passive behavioral sensing versus periodic clinical self-report has not been systematically decomposed, leaving it unclear whether behavioral data adds an independent predictive signal or primarily amplifies symptom history[23,24,26]. Third, the behavioral features associated with deteriorating depressive symptoms, including which patterns precede worsening and how consistent those patterns are across individuals, remain poorly characterized. There is growing evidence that behavioral features covary with mental health in highly person-specific ways, such that group-level findings systematically understate the magnitude and variability of individual-level associations[19,27–29]. However, the empirical basis confirming this heterogeneity remains limited.

We address these gaps through empirical analysis of data obtained in the Human Screenome Project, specifically assessments of depressive symptoms (CES-D)[30] obtained every two weeks alongside behavioral features derived from over 100 million screenshots (5-second resolution) obtained over one year from 96 adults. We derived behavioral features across five conceptual domains, each grounded in prior work[6,10,19,31–33] linking smartphone behavior to mental health: dosage (overall device use intensity), fragmentation (structure of device engagement), circadian pattern (timing of device use), social media (platform-specific engagement), and content diversity (variety of visual content consumed). We use these features to predict clinically meaningful changes in CES-D scores over the following two-week interval, classifying each fortnightly period as worsening, remaining stable, or improving (referred to throughout as trajectory direction prediction) based on three operationalizations: crossing absolute clinical severity thresholds, exceeding each person's own historical variability, and dividing the change distribution into equal-sized groups. We evaluate the quality of prediction using temporal holdout within-person as the primary design and leave-group-out cross-validation to assess generalization to new individuals with no prior assessment history. We additionally characterize how performance varies when the most recent assessment is several weeks old and across participants with different symptom patterns, and systematically compare behavioral features against prior symptom history to isolate what passive sensing independently contributes.

# Results

## *Sample Characteristics*

Ninety-six adults from the Human Screenome Project participated in this study, completing between 10 and 25 fortnightly CES-D surveys over approximately one year (M = 20.9, SD = 3.9, Median = 22; total of 2,002 person-period observations). Demographic characteristics are presented in Table 1.

**Table 1 | Sample demographic characteristics (N = 96).** Values are N (%) or mean (SD). Race was assessed as select all that apply; percentages sum to more than 100%. Participants counted in the Multiracial row are also counted within each of the individual race categories they selected.

| **Characteristic** | **N (%)** |
|---|---|
| **Age, years** | M = 47.4 (SD = 15.7); range 20–78 |
| **Gender** | |
| Female | 56 (58.3) |
| Male | 38 (39.6) |
| Other | 2 (2.1) |
| **Race** | |
| White | 72 (75.0) |
| Black or African American | 16 (16.7) |
| Asian | 7 (7.3) |
| Native American or Alaska Native | 6 (6.2) |
| Pacific Islander | 1 (1.0) |
| Other | 10 (10.4) |
| Multiracial (2 or more selected) | 11 (11.5) |
| **Ethnicity** | |
| Hispanic or Latinx | 16 (16.7) |
| **Education** | |
| Less than high school | 1 (1.0) |
| High school diploma or GED | 17 (17.7) |
| Some college | 27 (28.1) |
| 2-year college degree | 15 (15.6) |
| 4-year college degree | 28 (29.2) |
| Master's degree | 7 (7.3) |
| Professional degree (JD/MD) | 1 (1.0) |
| **Household income** | |
| Below $25,000 | 38 (39.6) |
| $25,000–$99,999 | 22 (22.9) |
| $100,000 or above | 30 (31.2) |
| Prefer not to answer or don't know | 6 (6.3) |
| **Marital status** | |
| Married or living as married | 46 (47.9) |
| Single/never married | 28 (29.2) |
| Divorced or separated | 17 (17.7) |
| Widowed | 5 (5.2) |

*CES-D trajectories reveal substantial within-person variability.*

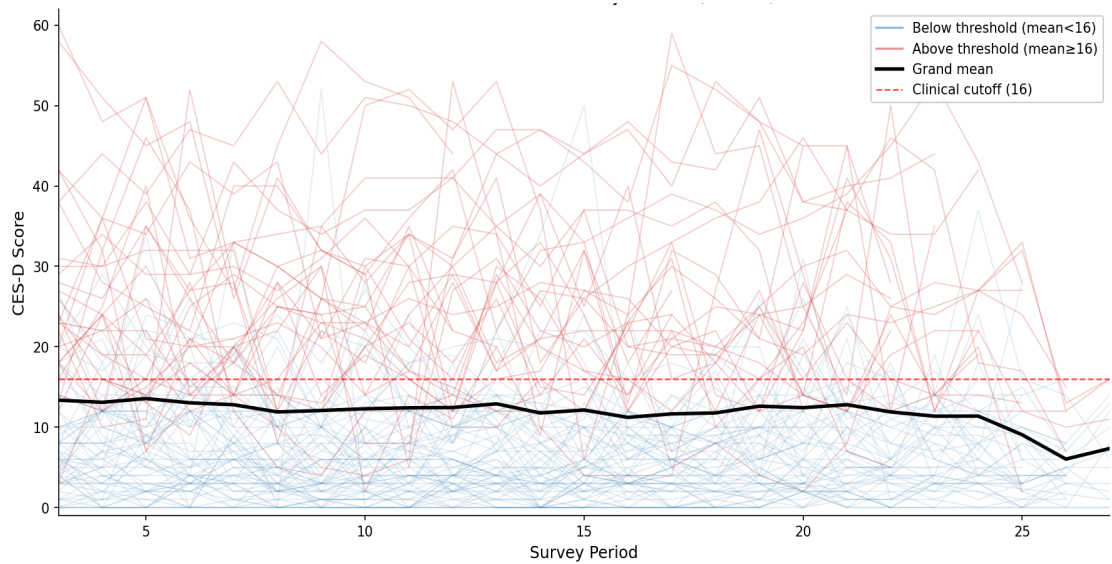

**Figure 1 | Within-person CES-D trajectories over one year (N = 96).** Each line represents how one participant's CES-D score changed across fortnightly assessment periods (2,002 periods total across all participants). Lines are colored by whether participants' mean CES-D score was above (red) or below (blue) the clinical threshold of 16. The red dashed horizontal line indicates the clinical threshold. The grand mean trajectory across all participants is shown in black. The vertical spread between lines reflects stable between-person differences in typical symptom level (ICC = 0.763, where ICC represents the proportion of total variance that is between people). The fluctuations within each line reflect within-person variability over time (23.7% of variance in raw CES-D scores; 99.6% of variance in fortnightly CES-D change). CES-D, Center for Epidemiologic Studies Depression Scale.

CES-D scores varied substantially both across and within individuals over the study year (Figure 1). To quantify this variation, we computed the intraclass correlation coefficient (ICC), which represents the proportion of total variance that is between people, for both CES-D scores and fortnightly CES-D change scores (computed as the difference between consecutive fortnightly assessments).

For raw CES-D scores, 76.3% of variance was between-person (ICC = 0.763), reflecting stable individual differences in typical symptom level, visible as the vertical spread of lines in Figure 1. The remaining 23.7% of variance was within-person, representing change over time around each individual's baseline. For fortnightly CES-D change, only 0.4% of variance was between-person (ICC = 0.004), with 99.6% within-person, visible as fluctuations within each trajectory. This indicates that changes are highly variable within individuals across time rather than following stable upward or downward trends.

Symptom fluctuation across the study year was clinically meaningful in scale. Of the 96 participants, 63.5% (N=61) had scores at or above the clinical threshold of 16 on one or more occasions during the study, and 55.2% (N=53) of the trajectories crossed the clinical threshold two or more times in either direction. That is, most participants did not occupy a stable clinical category during the study but moved in and out of clinical risk repeatedly. Looking specifically at point changes between assessments, 84.4% (N=81) experienced at least one change of five or more points between consecutive assessments. The extent of within-person fluctuation, as indicated by the within-person standard deviation of raw CES-D scores (iSD) across the year differed substantially across persons (Mean = 5.11, range = 0.50 to 17.86). Some participants exhibited minimal fluctuation, while others experienced large swings over the year. These patterns confirm that most individuals experienced substantial changes in CES-D over the study year, supporting the need for within-person prediction models that account for both stable individual differences in baseline severity and dynamic, time-varying changes.

### *Within-person CES-D trajectory direction is predictable across operationalizations and models.*

CES-D trajectories show substantial within-person variability, with most participants crossing clinical thresholds multiple times over the study year. Whether that variability is predictable remains the central question: can behavioral and symptom history forecast whether a given individual's CES-D score will worsen, remain stable, or improve over the following fortnight? We address this by training four classification models (XGBoost, LightGBM, SVM, and ElasticNet) using the first 60% of each participant's time series for training, the next 20% for hyperparameter selection, and the final 20% as a held-out test set, evaluated using AUC (one-vs-rest macro-average), balanced accuracy, sensitivity for the worsening class, and PPV (positive predictive value for the worsening class).

Three operationalizations of this outcome were evaluated: severity crossing (primary), which defines these classes based on movement across established CES-D clinical thresholds[30,34,35]; personalized threshold, which defines them relative to each person's own historical variability; and balanced tercile, which enforces equal class proportions to assess performance under balanced priors. Results for severity crossing and personalized threshold are presented in Tables 2 and 3; balanced tercile results are reported in Supplementary Materials. Full modeling details are provided in Methods.

**Table 2** | Classification performance, severity crossing label (held-out test set, N = 411; 9% worsening, 11% improving, 80% stable)

|  | Balanced accuracy | AUC (OvR) | Sensitivity, worsening | PPV, worsening |
|---|---|---|---|---|
| XGBoost<br>Best AUC | 0.834<br>(0.784, 0.882) | 0.906<br>(0.881, 0.929) | 0.838<br>(0.707, 0.947) | 0.356<br>(0.253, 0.464) |
| LightGBM<br>Best Accuracy and Sensitivity | 0.842<br>(0.798, 0.890) | 0.901<br>(0.876, 0.925) | 0.865<br>(0.744, 0.970) | 0.344<br>(0.247, 0.439) |
| SVM | 0.696<br>(0.628, 0.766) | 0.841<br>(0.806, 0.876) | 0.649<br>(0.488, 0.806) | 0.304<br>(0.203, 0.405) |
| ElasticNet | 0.699<br>(0.626, 0.764) | 0.834<br>(0.797, 0.871) | 0.730<br>(0.576, 0.871) | 0.245<br>(0.167, 0.336) |
| Rule-based baselines |  |  |  |  |
| Regression to person's mean<br>Strongest baseline | 0.674 | 0.750 | 0.541 | 0.408 |
| Last value carried forward | 0.336 | 0.557 | 0.054 | 0.053 |
| Person-specific modal class | 0.327 | 0.499 | 0.000 | — |
| Predict all stable | 0.333 | 0.500 | 0.000 | — |

Values for machine learning models are point estimates with 95% bootstrap confidence intervals (1,000 resamples, seed = 42) shown below each estimate in smaller text. Baseline values are point estimates without confidence intervals. Green shading indicates the model with the highest AUC. Grey shading indicates rule-based baselines. — indicates undefined (positive predictive value is undefined when no worsening predictions are made). Balanced accuracy, average sensitivity across all three trajectory classes (worsening, stable, improving). AUC, area under the receiver operating characteristic curve, one-vs-rest macro-average. Sensitivity (worsening), proportion of true worsening episodes correctly identified. PPV (worsening), proportion of worsening predictions that were correct. The regression-to-person-mean baseline predicts the direction in which each person's current CES-D score would need to move to return to their typical severity band. OvR, one-vs-rest; PPV, positive predictive value; CES-D, Center for Epidemiologic Studies Depression Scale.

Under the severity crossing operationalization of CES-D change, XGBoost achieved the highest AUC at 0.906 (0.881, 0.929) and balanced accuracy of 0.834, correctly identifying 31 of 37 true worsening episodes in the held-out test set. LightGBM achieved the highest balanced accuracy (0.842) and sensitivity for detecting worsening episodes (worsening sensitivity) of 0.865, catching 32 of 37 worsening episodes, with an AUC of 0.901 (0.876, 0.925). Both tree-based models substantially outperformed the linear models, where ElasticNet achieved an AUC of 0.834 and SVM achieved an AUC of 0.841. Positive predictive value for the worsening class was 0.356 for XGBoost and 0.344 for LightGBM, corresponding to approximately 2 false alarms per true worsening detection. The practical implications of this trade-off are discussed in the context of prediction model deployment later in the Discussion.

As seen in Table 2, all four prediction models substantially exceeded all rule-based baselines. Predicting stable for every observation or predicting each person's most frequent historical class yielded zero worsening detections. The regression-to-person-mean baseline, which predicts trajectory direction based on whether a person's current CES-D score is above or below their own typical level without behavioral features or learned parameters, was the strongest rule-based comparator with an AUC of 0.750 and worsening sensitivity of 0.541. The full XGBoost model exceeded this by 0.156 AUC points and detected 11 additional worsening episodes (31 vs 20 of 37).

**Table 3** | Classification performance, personalized threshold label (held-out test set, N = 411; 10% worsening, 12% improving, 78% stable)

|  | Balanced accuracy | AUC (OvR) | Sensitivity, worsening | PPV, worsening |
|---|---|---|---|---|
| XGBoost | 0.620 (0.551, 0.682) | 0.750 (0.707, 0.793) | 0.634 (0.485, 0.788) | 0.166 (0.105, 0.224) |
| LightGBM | 0.586 (0.524, 0.654) | 0.731 (0.689, 0.780) | 0.463 (0.341, 0.585) | 0.145 (0.091, 0.204) |
| SVM | 0.620 (0.550, 0.686) | 0.751 (0.702, 0.796) | 0.610 (0.454, 0.771) | 0.191 (0.126, 0.257) |
| ElasticNet *Best AUC* | 0.603 (0.534, 0.669) | 0.755 (0.704, 0.802) | 0.561 (0.400, 0.724) | 0.175 (0.110, 0.236) |
| *Rule-based baselines* | | | | |
| Regression to person's mean *Strongest baseline* | 0.533 | 0.642 | 0.293 | 0.240 |
| Last value carried forward | 0.322 | 0.529 | 0.049 | 0.045 |
| Person-specific modal class | 0.333 | 0.500 | 0.000 | — |
| Predict all stable | 0.333 | 0.500 | 0.000 | — |

Values for machine learning models are point estimates with 95% bootstrap confidence intervals (1,000 resamples, seed = 42) shown below each estimate in smaller text. Baseline values are point estimates without confidence intervals. Green shading indicates the model with the highest AUC. Grey shading indicates rule-based baselines. — indicates undefined (positive predictive value is undefined when no worsening predictions are made). Balanced accuracy, average sensitivity across all three trajectory classes (worsening, stable, improving). AUC, area under the receiver operating characteristic curve, one-vs-rest macro-average. Sensitivity (worsening), proportion of true worsening episodes correctly identified. PPV (worsening), proportion of worsening predictions that were correct. The regression-to-person-mean baseline predicts the direction in which each person's current CES-D score would need to move to return to their typical severity band. OvR, one-vs-rest; PPV, positive predictive value; CES-D, Center for Epidemiologic Studies Depression Scale.

Table 3 shows results under the personalized threshold operationalization, where the prediction task is harder: worsening is defined relative to each person's own historical variability rather than fixed clinical thresholds, removing the boundary proximity effect that favored tree-based models under severity crossing. Under this operationalization, linear models were competitive with tree-based models, with ElasticNet achieving the highest AUC at 0.755 (0.704, 0.802) and XGBoost the highest worsening sensitivity at 0.634 (0.485, 0.788), catching 26 of 41 true worsening episodes. As with severity crossing, predicting stable for every observation or predicting each person's most frequent historical class yielded zero worsening detections. The regression-to-person-mean baseline was again the strongest rule-based comparator at AUC 0.642, and all four models exceeded it.

*Each person's typical symptom baseline strongly boosts predictive power.*

Having established that within-person trajectory direction is predictable and that learned models substantially exceed rule-based alternatives, we next ask: what drives that predictive performance? To decompose the relative contribution of each data source, we conducted a nested feature ablation, systematically adding feature sets one at a time and testing whether each addition produces a statistically significant improvement in AUC.

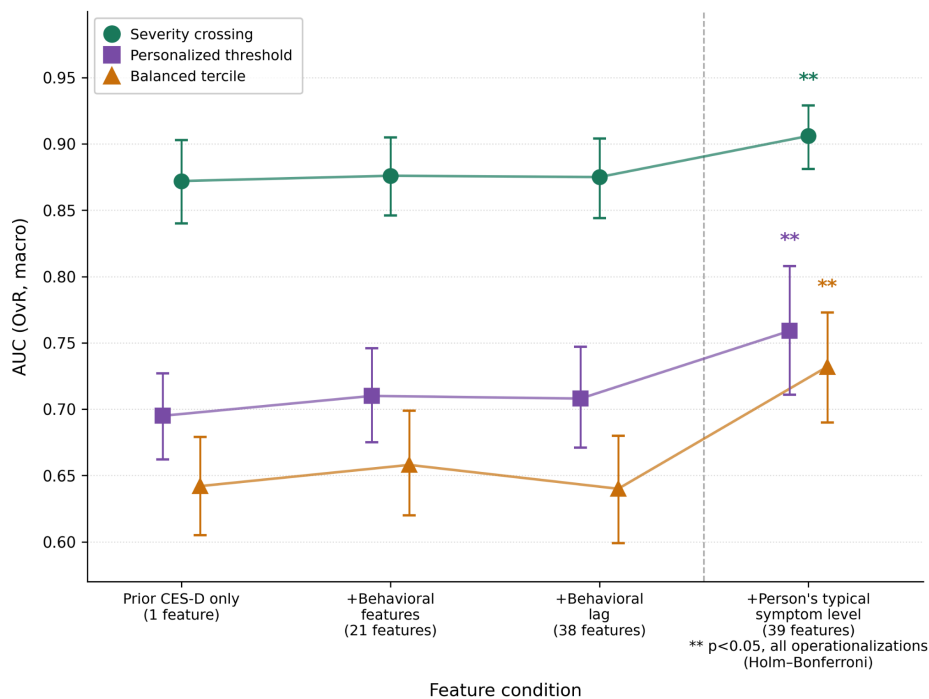

**Figure 2 | Feature ablation across label operationalizations.** AUC (one-vs-rest macro-average) at each of four nested feature conditions, shown for the best-performing model per label operationalization: XGBoost for severity crossing, ElasticNet for personalized threshold, and LightGBM for balanced tercile. The four conditions are: (1) most recently available CES-D score only, (2) adding current-period behavioral features, (3) adding behavioral lag features from the prior period, and (4) adding each person's typical symptom level. Each condition adds cumulatively to the prior. Neither step 2 (behavioral features vs. step 1) nor step 3 (behavioral lag vs. step 2) produces a statistically significant gain over the preceding step. Adding each person's typical symptom level (step 4) is the only significant step, surviving Holm-Bonferroni correction across all three label operationalizations. Error bars show 95% bootstrap confidence intervals (1,000 resamples). **, $p < 0.05$, Holm-Bonferroni corrected. AUC, area under the receiver operating characteristic curve; OvR, one-vs-rest; CES-D, Center for Epidemiologic Studies Depression Scale.

Figure 2 shows AUC at each feature condition for the best-performing model per label operationalization. Under the severity crossing label, the most recently available CES-D score alone achieves AUC = 0.872. This strong starting point reflects the predictive power of periodic self-report: although only a single prior CES-D value enters as a feature at each prediction point, the model learns from the full sequence of such observations across training, capturing how current symptom position relates to subsequent trajectory direction. In particular, proximity to a severity boundary is itself highly informative about the likely direction of the next assessment. Adding the full suite of Screenome-derived behavioral features and their temporal lags yields no statistically significant improvement: the gains at step 2 (behavioral features, +0.004) and step 3 (behavioral lag, -0.002) are non-significant across all three label operationalizations (all $p > 0.05$, Holm-Bonferroni corrected). The only statistically significant step is the addition of each person's typical symptom level. AUC increases by 0.031 ($p = 0.018$, Holm-Bonferroni corrected), with the gain significant across all three label operationalizations after Holm-Bonferroni correction. A two-feature model using only the most recently available CES-D score and each person's typical symptom level matches the full 39-feature XGBoost model across all metrics (AUC = 0.917 vs AUC = 0.906 respectively, overlapping confidence intervals; 99.0% prediction agreement; Supplementary Table S5), confirming that behavioral features do not add discriminative power beyond these two clinical anchors.

The clinical significance of this finding is illustrated by a specific failure mode. Without each person's typical symptom level, the model detects 0% of moderate-to-severe worsening transitions, the most consequential subtype. With it, detection rises to 88% (7 of 8 cases). A current CES-D score of 20 carries a different clinical meaning for a person who typically scores 8 versus one who typically scores 22, and without the baseline anchor, the model cannot make this distinction. The gain from a person's typical symptom level grows as the clinical boundary proximity effect is removed across label operationalizations, from +0.031 under severity crossing to +0.052 under personalized threshold and +0.091 under balanced tercile, confirming that this feature encodes genuine chronic trait information rather than exploiting the severity boundary structure. Full AUC results across all four models and all three label operationalizations at each feature condition are reported in Supplementary Table S4.

Although behavioral features do not independently add predictive power above the CES-D anchors, they characterize how deterioration manifests in digital behavior in the fortnightly periods preceding threshold crossings. These associations are examined in the following section.

### *Behavioral features show small but theoretically consistent prodromal associations.*

The preceding section established that Screenome-derived behavioral features do not independently add statistically significant predictive power above the CES-D anchors. The question examined here is distinct: what behavioral information does the model engage, and do the features it selects correspond to theoretically meaningful signals of deterioration? We address this using two complementary evidence sources. XGBoost feature importance provides evidence of feature engagement from the best-performing model across all trajectory classes (Figure 3). ElasticNet signed coefficients characterize the direction of behavioral associations with worsening (Figure 4).

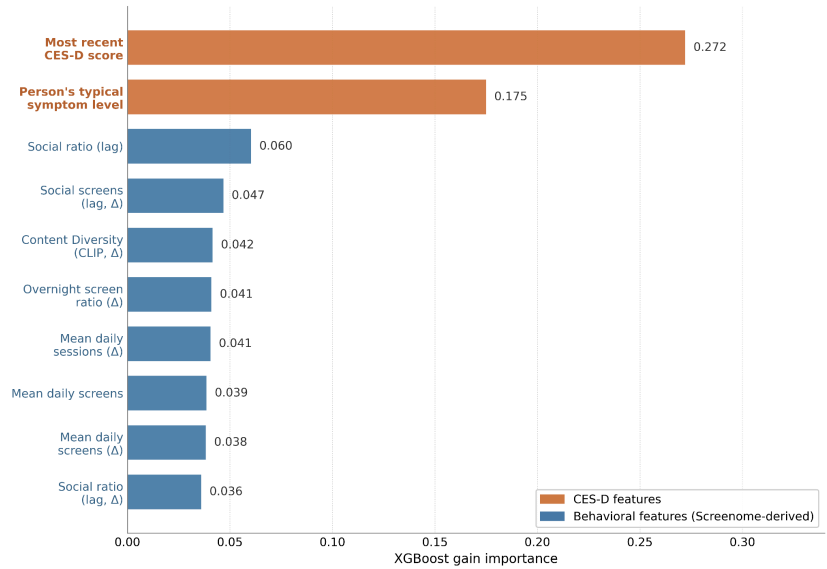

**Figure 3 | XGBoost feature importance, severity crossing label, top 10 features.** Gain importance reflects the average reduction in prediction loss at each split using each feature, aggregated across all trees and all trajectory classes (worsening, stable, improving). Shown are the top 10 of 18 features with nonzero gain out of 39 total. CES-D features account for 45% of the total gain. The eight behavioral features in the top ten span all five conceptual domains defined in the methods: social media engagement, circadian pattern, fragmentation, content diversity, and dosage. Gain importance is unsigned and cross-class; class-specific directional associations with worsening are shown in Figure 4. Δ denotes period-over-period change from the preceding fortnightly window.

Figure 3 shows XGBoost feature importance for the top ten features under severity crossing. The two CES-D anchor features together account for 45% of the total gain. The remaining top-ranked features span all five conceptual feature domains defined in the methods: dosage, fragmentation, circadian pattern, social media engagement, and content diversity. Their presence among the top-ranked features confirms that behavioral features are genuinely engaged by the model across trajectory classes, and that the features engaged correspond to domains with established theoretical links to depressive symptom change.

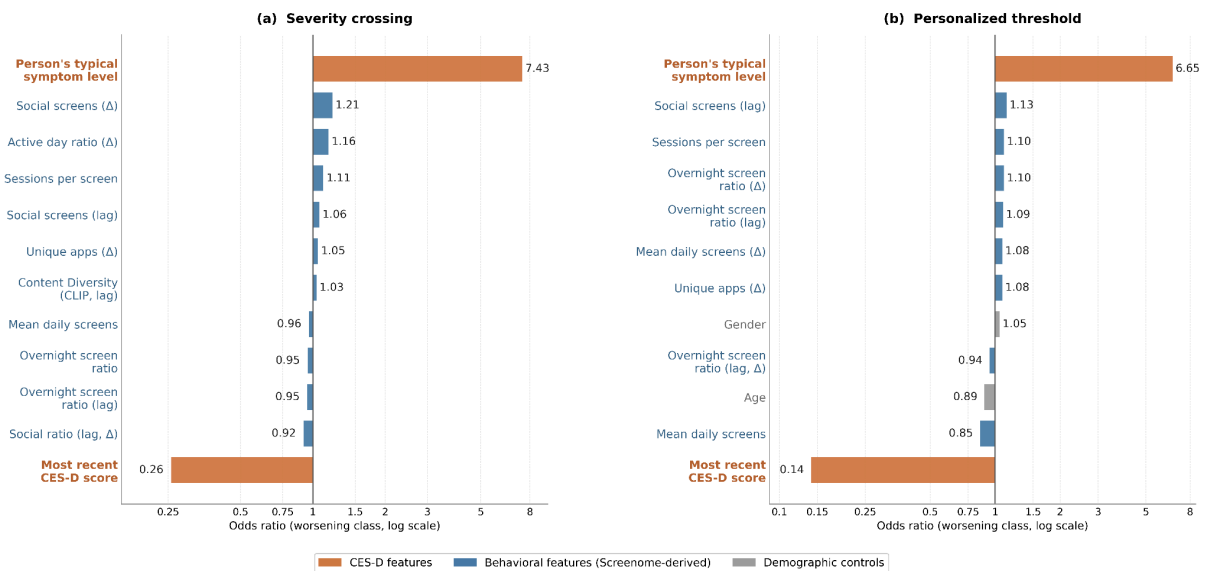

**Figure 4 | ElasticNet worsening-class odds ratios, severity crossing and personalized threshold.** (a) Severity crossing. (b) Personalized threshold. Both panels display the coefficients of the grid-searched ElasticNet models listed in Supplementary Table S8, fitted on z-score standardized features; top 12 features by absolute coefficient magnitude shown. Odds ratios represent the change in worsening odds per one standard deviation increase in each feature, derived by exponentiating the standardized ElasticNet coefficients. OR > 1 indicates increased worsening probability; OR < 1 indicates decreased probability. The reference line at OR = 1 indicates no association. X-axis is on a logarithmic scale; x-axis ranges differ between panels. Δ denotes period-over-period change from the preceding fortnightly window; CES-D, Center for Epidemiologic Studies Depression Scale.

Figure 4 shows ElasticNet worsening-class odds ratios under both label operationalizations. Across both labels, a person's typical symptom level carries the strongest positive association with worsening (OR = 7.43 under severity crossing, OR = 6.65 under personalized threshold). The most recently available CES-D score is negatively associated with worsening (OR = 0.26 and 0.14 respectively): a currently elevated score, in the context of a person's own baseline, predicts improvement rather than further escalation.

Among behavioral features, several patterns emerge. Escalating social media engagement is associated with worsening: increasing social screen time under severity crossing (OR = 1.21, Figure 4a), and lagged social screen level under both labels (OR = 1.06 under severity crossing; OR = 1.13 under personalized threshold). Reduced overall engagement accompanies this: mean daily screens is inversely associated under both labels, more strongly under personalized threshold (OR = 0.96 under severity crossing; OR = 0.85 under personalized threshold). Active day ratio change is positively associated with worsening under severity crossing (OR = 1.16, Figure 4a). Behavioral fragmentation is reflected in higher session rate (sessions per screen, OR = 1.11 under severity crossing; OR = 1.10 under personalized threshold) and increasing app diversity (OR = 1.05 under severity crossing; OR = 1.08 under personalized threshold). Overnight screen use shows divergent directions across operationalizations: under severity crossing, current-period and lagged overnight ratio levels are both inversely associated with worsening (OR = 0.95 for each, Figure 4a), while under personalized threshold, three temporally distinct indicators emerge: sustained prior-period overnight use (OR = 1.09), current-period escalation (OR = 1.10), and a negative lagged change coefficient (OR = 0.94, Figure 4b). Demographic covariates were also selected under personalized threshold: female gender was positively associated (OR = 1.05), and older age was inversely associated with worsening (OR = 0.89).

In stark contrast to the CES-D anchors, which carry odds ratios far from unity (ranging from 0.14 to 7.43 across panels), the behavioral and demographic odds ratios cluster close to 1, ranging from 0.92 to 1.21 under severity crossing and 0.85 to 1.13 under personalized threshold. Figure 5 shows why. Within-person correlations between each behavioral feature's period-over-period change and CES-D change span the full range from approximately -0.6 to +0.75 across the 96 participants, with group means near zero for every feature (range -0.011 to +0.057) and standard deviations of 0.249 to 0.286. For the change in social screen time, 57% of participants show positive within-person correlations, meaning escalating social screen use co-occurs with CES-D increases for them, while 43% of participants show the opposite. Mean absolute within-person correlations of 0.20-0.24 confirm that individual-level associations are real; they cancel at the group level because the same behavioral change is associated with rising CES-D for some individuals and falling CES-D for others. The group-level associations in Figure 4 thus conceal substantial individual variation, which is why the model relies primarily on CES-D trajectory features for prediction. We return to this in the discussion.

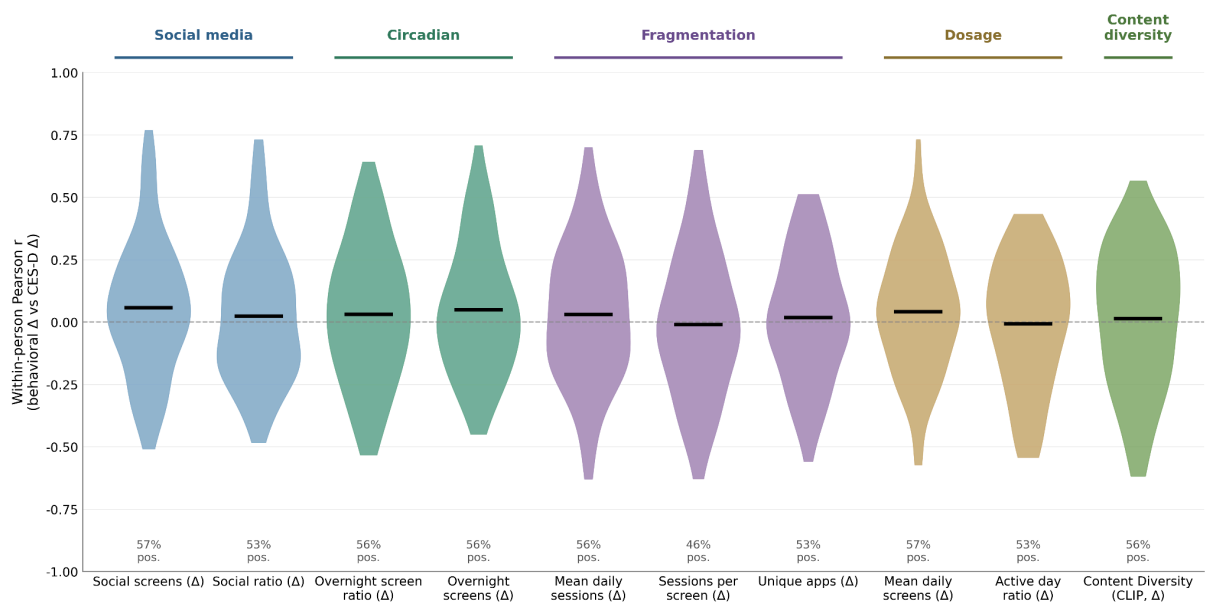

**Figure 5 | Within-person correlations between behavioral feature changes and CES-D change (N = 96 participants).** Each violin shows the distribution of within-person Pearson correlations between a behavioral feature's period-over-period change and CES-D change across all 96 participants. The horizontal bar indicates the group mean. Participants are excluded per feature if their within-person variance for that feature is zero (N ranges from 59 to 96 per feature). A positive correlation indicates that increases in the behavioral feature co-occur with increases in CES-D score within that person; a negative correlation indicates the opposite. The dashed reference line at r = 0 indicates no within-person association. Features are grouped by conceptual domain. Percentage positive (% pos.) indicates the proportion of participants with a positive within-person correlation for that feature. Δ denotes period-over-period change from the preceding fortnightly window.

### *The model generalizes to new individuals, degrades gracefully with stale assessments, and shows heterogeneous performance across participants.*

Having shown that within-person trajectory direction is predictable, that prediction depends primarily on each person's CES-D anchors, and that behavioral features add mechanistic insight rather than independent predictive power, we next ask how the model performs when the inputs it relies on are partial or absent. We tested two main scenarios a monitoring system will inevitably encounter: a new individual presenting for their first assessment with no accumulated history, evaluated using leave-group-out cross-validation; and a returning individual whose most recent CES-D score is several weeks old, evaluated by replacing the current assessment with a stale one at prediction time. Results are reported in Table 4.

**Table 4 |** Predictive performance under deployment scenarios, severity crossing label (XGBoost, held-out test set, N = 411 unless noted)

|  | **Balanced accuracy** | **AUC (OvR)** | **Sensitivity, worsening** | **PPV, worsening** |
|---|---|---|---|---|
| **Full model** Known person, current assessment | 0.834 (0.784, 0.882) | 0.906 (0.881, 0.929) | 0.838 (0.707, 0.947) | 0.356 (0.253, 0.464) |
| **Cold start** New person, current assessment | 0.720 (0.693, 0.744) [a] | 0.821 (0.808, 0.832) [a] | 0.569 (0.436, 0.636) [a] | 0.224 (0.182, 0.238) [a] |
| **Stale assessment, 4 weeks**<br>Known person, 4-week-old CES-D | 0.565<br>(0.489, 0.644) | 0.735<br>(0.682, 0.786) | 0.514<br>(0.333, 0.667) | 0.232<br>(0.143, 0.321) |
| **Stale assessment, 8 weeks**<br>Known person, 8-week-old CES-D | 0.507<br>(0.434, 0.581) | 0.702<br>(0.647, 0.754) | 0.432<br>(0.276, 0.607) | 0.188<br>(0.106, 0.276) |

Values are point estimates with 95% bootstrap confidence intervals (1,000 resamples, seed = 42) shown in parentheses. Green shading indicates the reference condition (full model). All scenarios use the 39-feature XGBoost model trained on the full training set; rows differ only in what information is available at prediction time. Stale assessment conditions replace the most recent CES-D with the participant's value from one (4 weeks) or two (8 weeks) prior assessment periods. [a] Cold-start values are reported as the mean across 5 leave-group-out cross-validation repeats (5 × 5 folds, approximately 19–20 participants per fold held out entirely from training, person's typical symptom level set to the population mean across training participants); the interval reports the minimum and maximum across the 5 repeats. AUC, area under the receiver operating characteristic curve, one-vs-rest macro-average. Balanced accuracy, average sensitivity across all three trajectory classes (worsening, stable, improving). Sensitivity (worsening), proportion of true worsening episodes correctly identified. PPV (worsening), proportion of worsening predictions that were correct. CES-D, Center for Epidemiologic Studies Depression Scale.

The first deployment question is whether the model generalizes to participants it has never seen during training: the scenario of onboarding a new participant. Under leave-group-out cross-validation, where approximately 19-20 participants per fold were held out entirely from training, cold-start AUC was 0.821 (0.808, 0.832), with worsening sensitivity of 0.569 and balanced accuracy of 0.720. The drop from the full model (AUC = 0.906) reflects the cost of losing each person's typical symptom level as a calibrated anchor, but cold-start performance remains well above all rule-based comparators, including the strongest non-model baseline of regression to person mean (AUC = 0.750, Table 2). These cold-start results indicate that the model generalizes to participants unseen during training, retaining approximately 91% of the full model's AUC despite losing the person-level CES-D anchors.

The second deployment question concerns assessment currency: what happens to predictive performance when a returning participant's most recent CES-D score is no longer current? With a four-week-old assessment, AUC dropped to 0.735 (0.682, 0.786), and worsening sensitivity fell to 0.514, compared to 0.906 and 0.838 respectively, under the full model with a current assessment. With an eight-week-old assessment, performance degraded further to AUC 0.702 (0.647, 0.754) and sensitivity 0.432, indicating that more than half of worsening episodes went undetected. The staleness gradient is steep: much of the performance loss occurs already at four weeks, with further deterioration at eight weeks. Notably, a known participant with an eight-week-old CES-D (AUC = 0.702) performs worse than a new participant presenting for the first time with a current assessment (AUC = 0.821).

These deployment characterizations describe average performance across the test sample. Figure 6 illustrates individual variation in prediction performance across the test sample. For participants whose CES-D trajectories span multiple severity zones with substantial fluctuation, the model assigns high worsening probability at the correct periods with few errors elsewhere. For participants whose trajectories remain in the minimal severity range, worsening events go largely undetected, with predicted worsening probability remaining flat even when these individuals cross a clinical threshold. Full confusion matrices across all trajectory classes are provided in Supplementary Table S2.

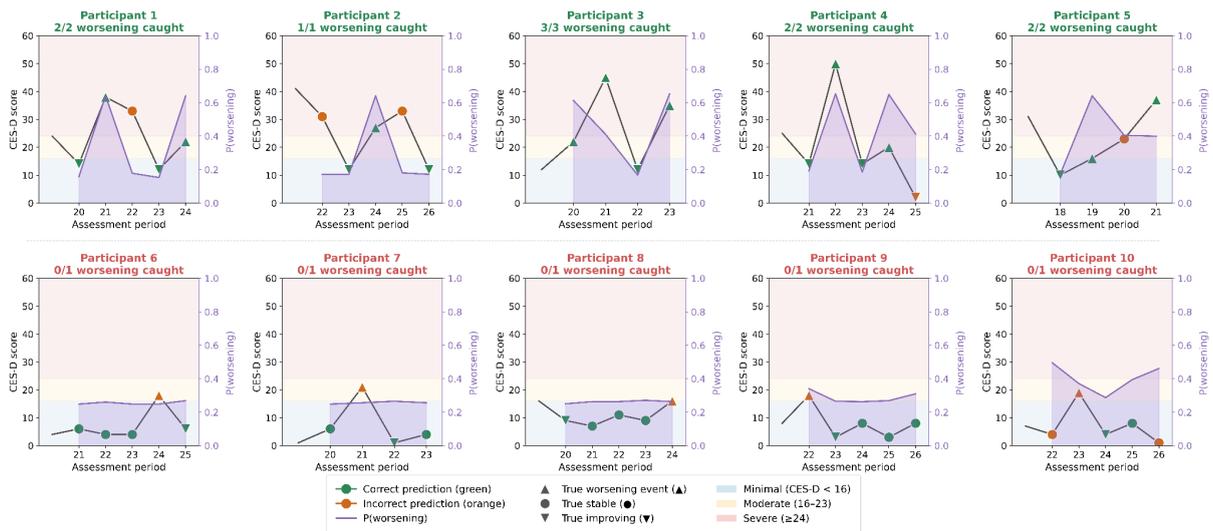

**Figure 6 | Sample individual CES-D trajectories and model worsening predictions from the test set, illustrating detection success and failure.** Top row: five sample participants for whom all worsening events were correctly identified; shown are those with the clearest contrast between predicted worsening probability at true worsening periods and non-worsening periods (highest ratio of mean P(worsening) at worsening versus non-worsening periods), among all test-set participants with perfect worsening sensitivity. Bottom row: five sample participants whose worsening events were all missed; shown are those with the lowest model confidence at true worsening periods among participants with zero worsening sensitivity. The grey line shows the observed CES-D trajectory. At each fortnightly period, the model receives the prior CES-D score and behavioral features and produces a predicted worsening probability, shown as the purple shaded line (P(worsening), right axis). The prediction is made at the start of the period for the end of the period. The colored dot shows the actual outcome at the end of that period, where the CES-D landed at the next assessment, sitting on the trajectory line at the resulting CES-D value. The dot shape indicates the true outcome class: upward triangle (▲) for worsening (crossing into a higher severity band), circle (●) for stable (remaining in the same band), downward triangle (▼) for improving (crossing into a lower band). Dot color indicates prediction accuracy: green for a correctly classified period, orange for an incorrect prediction regardless of which class was predicted. For example, an orange circle indicates the model predicted the wrong direction even though the true label was stable. Shaded background indicates CES-D severity zones: minimal (below 16), moderate (16–23), and severe (24 and above), following established CES-D clinical risk thresholds[30,34,35]. Full confusion matrices across all trajectory classes are provided in Supplementary Table S2.

## Discussion

In this paper, we found that trajectory direction is predictable (AUC = 0.906; worsening sensitivity 0.838) under temporal holdout and generalizes to unseen individuals (AUC = 0.821 under leave-group-out cross-validation). Each person's typical symptom level (mean CES-D across prior assessments) was the only statistically significant feature addition above the most recent symptom report, robust across all three label operationalizations and all four model families. Behavioral features did not, on their own, add predictive power beyond this clinical anchor. However, they revealed behavioral patterns preceding worsening, including increased social media engagement, reduced overall screen time with more fragmented device use, and changes in overnight activity, characterizing how deterioration manifests behaviorally before clinical thresholds are crossed. Within-person correlations between behavioral features and CES-D change were close to zero on average, not because behavioral associations were absent, but because the same pattern is associated with worsening for some individuals and improvement for others. These findings provide a proof-of-concept foundation for monitoring systems that could connect individuals to care before symptoms reach crisis.

*Within-person prediction as the appropriate paradigm*

Variance decomposition confirmed that 99.6% of the variance in fortnightly CES-D change was within-person rather than between-person (ICC = 0.004), indicating that changes are highly variable within individuals across time rather than following stable directional trends (Figure 1). This finding has important methodological implications: designs that compare individuals who differ in average symptom levels cannot recover the dynamics governing any individual's trajectory over time. More broadly, this reorients the prediction question itself: not where does this person stand relative to others, but where is this person headed relative to their own history.

The model's design reflects this within-person focus. Prediction is anchored to each person's own symptom baseline, their behavioral history is carried forward through lag features, and evaluation is strictly temporal within individuals. At approximately 20 fortnightly CES-D observations per person, the combination of limited individual time series length and the temporal holdout requirement leaves insufficient data for stable person-specific parameter estimation within a forecasting framework. Calibrating behavioral feature weights to each person's own historical associations with symptom change, within a forecasting design, represents a promising direction as longitudinal datasets scale, shifting from a single population-level model toward transferable models derived from empirically identified behavioral subgroups, where clustering is data-driven rather than defined by demographic or clinical categories a priori [19,36].

*The role of the symptom baseline and classification label in trajectory prediction*

Prediction using the most recent CES-D score alone achieves an AUC of 0.872 under temporal holdout, demonstrating that periodic digital self-report has genuine predictive value as a starting point for screening and monitoring programs that reach people outside the clinic. This reframes the contribution of passive behavioral sensing not as a replacement for clinical information that most people with depression never receive, but as augmentation of lightweight self-report that a digital health system can collect at low burden.

The addition of each person's typical symptom level as a second feature is the only statistically significant step in the feature ablation. A current CES-D score of 18 carries a different risk for someone who typically scores 8 than for someone who typically scores 22. Without the baseline anchor, the model cannot distinguish between acute escalation and chronic moderate depression. This is illustrated in moderate-to-severe worsening detection, which rises from 0% to 88% when each person's typical symptom level is added. This also explains the negative association of the most recently available CES-D score with worsening: a currently elevated score, in the context of a person's own baseline, signals that the person is above their typical level and therefore more likely to improve than to escalate further. Screening and monitoring programs must accumulate a per-person assessment history before the anchor is reliable, aligning with the broader principle that each person's own history, rather than population norms, is the appropriate reference frame for detecting clinically meaningful change[37]. Figure 6 illustrates what happens when this anchor provides insufficient signal. For participants whose trajectories span multiple severity zones, the model correctly identifies worsening periods with high confidence. For participants whose trajectories mostly remain "stable" in their severity range, worsening events go largely undetected: large sudden jumps from low baselines are underrepresented in training data, and prior CES-D scores provide little advance signal for this subgroup.

The two label operationalizations represent different clinical tasks. Under severity crossing, the CES-D anchor features are structurally aligned with the label definition; both are organized around clinical risk threshold boundaries, which means some of the model's strong performance reflects this alignment rather than purely its predictive power. Personalized threshold removes that alignment, requiring the model to detect change that is unusual for that specific individual regardless of absolute position. This is a harder task, reflected in the lower performance (AUC = 0.755 vs AUC = 0.906), but arguably a more clinically meaningful one for individuals whose symptoms persistently remain above or below clinical thresholds, for whom absolute cutoffs are less informative than deviation from their own typical level. Prospective validation in a diagnosed clinical sample is needed to determine which operationalization better predicts clinically meaningful deterioration.

*What behavioral features contribute and what they reveal*

The finding that Screenome-derived behavioral features across five conceptual domains (dosage, fragmentation, circadian pattern, social media engagement, and content diversity) do not independently add predictive power above periodic self-report (Figure 2) is itself informative. Prior digital phenotyping work has rarely decomposed these contributions using a systematic feature ablation, leaving it unclear whether passive sensing adds signal beyond what symptom history alone provides. Behavioral features address a complementary question to the CES-D anchors: not who will worsen and when, but how deterioration unfolds in the weeks before threshold crossings occur. In this study, their value lies not in independent predictive power but in mechanistic insight, characterizing the behavioral process of worsening that would otherwise go unobserved between self-report assessments, and advancing understanding of how screen-based behavior relates to mental health dynamics, a question central to both digital phenotyping research and media effects research.

The convergence across methods points to three behavioral patterns associated with worsening: escalating social media screen time, reduced overall screen time with more fragmented device use, and changes in overnight screen use. These replicate patterns reported in prior work linking social media escalation[32], behavioral fragmentation[19], and late-night phone use[31] to depressive symptom worsening. Figure 4 shows consistent directional signals for social media engagement and device use fragmentation across both label operationalizations: rising social media time accompanied by more fragmented use, indicating quick checking behavior and rapid switching between screen and real-world. Overnight screen use and overall device engagement show direction-dependent patterns across the two operationalizations. Under personalized threshold, elevated and rising overnight use is positively associated with worsening, consistent with prior work. Under severity crossing, higher overnight use is instead associated with lower odds of crossing clinical thresholds. One interpretation is that each label captures a different kind of change: movement across population-defined clinical cutoffs versus deviation from a person's own typical pattern. Demographic associations were also observed under personalized threshold, with female gender positively associated and older age inversely associated with worsening, consistent with established gender differences in depressive symptoms [38] and age-related differences in smartphone use patterns[39].

The behavioral signals identified here are consistent with prior work, but their interpretation from behavioral metadata alone remains open. Both escalating social media screen time and increasing fragmentation could reflect symptom manifestation, as behavioral disruption is an established feature of depressive episodes[2,3], or regulatory behavior where individuals seek connection, manage negative affect, or avoid sustained engagement as distress rises[33]. Social media escalation additionally admits a causal interpretation, where exposure itself contributes to worsening mood[27,29]. These interpretations carry different theoretical implications for understanding the relationship between screen behavior and mental health dynamics. The retention of visual content diversity from CLIP embeddings among the model's selected features, despite its small coefficient, suggests the visual character of content consumption carries information beyond dosage and fragmentation alone. Content-level analysis of what individuals are actually watching and engaging with[40] before worsening is a promising direction the Screenome's screen-by-screen capture enables but was not examined here.

The modest magnitude of group-level behavioral associations does not indicate a lack of signal. Behavioral associations with worsening exist but are highly person-specific (Figure 5): within-person correlations between behavioral feature changes and CES-D change span from approximately -0.6 to +0.75 across participants, with near-zero group means for every feature. For social screen time, 57% of participants show positive within-person correlations while 43% show the opposite, yet mean absolute correlations of 0.20 to 0.24 confirm that individual-level associations are real. The same behavioral change predicts worsening for some individuals and improvement for others, explaining why group-level passive sensing studies consistently find small effects. Prior work has documented behavioral heterogeneity in cross-sectional and concurrent associations with depression severity, but its presence in a prospective prediction study using within-person evaluation and longitudinal behavioral features, quantified across a year-long sample, provides stronger and more direct evidence that individual-level behavioral patterns of deterioration cannot be fully recovered from group-level models alone.

*Implications for digital health and clinical deployment*

The community sample in this study, in which 63.5% of participants scored above the clinical CES-D threshold at least once, and 55.2% crossed it repeatedly, reflects a population occupying the gap between untreated distress and formal clinical care. Depression monitoring of this kind could serve two complementary clinical roles. For individuals not currently in care, it could flag those who might benefit from a clinical contact they would not otherwise initiate, before symptoms reach crisis. For individuals already seeing providers episodically, it could provide continuous behavioral context between appointments, giving clinicians richer information for clinical conversations and an earlier signal that care may need to occur sooner than the next scheduled contact. The model's performance and generalization properties are encouraging for more research toward potential deployment. It achieves an AUC of 0.906 with a worsening sensitivity of 0.838 in the held-out test set, generalizes to completely new individuals from their first assessment (AUC = 0.821), degrading gradually rather than catastrophically when assessments lapse.

Assessment currency is the dominant deployment variable (Table 4). A single intake CES-D combined with behavioral data achieves an AUC of 0.821 under leave-group-out cross-validation, remaining well above the strongest rule-based comparator (regression to person mean, AUC = 0.750), and enabling monitoring from the very first interaction without any accumulated history. A new person with a current assessment outperforms a known person with an eight-week-old assessment (AUC = 0.821 vs. AUC = 0.702), indicating that keeping periodic self-report current matters more than accumulating long histories with outdated scores. The staleness gradient suggests fortnightly collection is preferable to longer intervals where feasible.

At the model's default threshold, worsening sensitivity of 0.838 is paired with a positive predictive value of 0.356, translating to approximately two false alarms per true worsening detection. Whether this trade-off is acceptable depends on the action a flag triggers: a low-burden automated check-in or a brief self-reflection prompt tolerates a higher false alarm rate, while direct clinical contact warrants raising the decision threshold to accept reduced sensitivity in exchange for higher precision. The model outputs calibrated three-class probabilities, making this trade-off adjustable without retraining.

Not all settings will support continuous behavioral sensing at full Screenome granularity, but a tiered approach can still provide value. Fortnightly self-report alone achieves AUC 0.872 from a single prior symptom score collected regularly, requiring only a periodic digital survey and no passive sensing infrastructure (Figure 2). Metadata-level passive sensing, including app usage logs, session counts, and overnight activity derived from operating system events rather than screen capture, adds behavioral context at substantially lower storage and privacy burden. Full Screenome-level capture represents the richest tier, enabling content-level analysis of what individuals are actually viewing and engaging with, and is appropriate where infrastructure and participant consent support it [18]. This tiered structure means that meaningful monitoring of depressive symptom trajectories does not require rich behavioral sensing infrastructure from the outset, and can become progressively richer as passive sensing capacity allows, a property relevant to the broader scalability of digital phenotyping approaches for mental health monitoring.

*Limitations and future directions*

These findings represent proof-of-concept evidence from a single community sample recruited through the Qualtrics research panel and restricted to Android smartphone users (a requirement of the Screenomics capture application at the time of study). The study demonstrates that within-person CES-D trajectory direction is predictable, characterizes the prodromal behavioral patterns that accompany deterioration, and establishes generalization properties under simulated deployment conditions, but does not constitute clinical validation in patient populations or evidence that acting on model predictions would improve patient outcomes. Replication across independent cohorts with diverse demographics, clinical populations, and measurement contexts is the next necessary step before the approach can be recommended for clinical use.

Participants were recruited during the COVID-19 pandemic between June 2020 and November 2021. Behavioral patterns during this period, including reduced mobility, altered social routines, and elevated baseline distress, may not generalize to typical conditions, and the social media escalation pattern in particular may partly reflect pandemic-specific behavioral changes rather than depression-specific prodromal signals.

The CES-D is a widely used and validated self-report measure for depressive symptoms, but as a sum score, it aggregates heterogeneous symptoms that may obscure differences in individual symptom trajectories, etiologies, and clinical relevance[3,35], a limitation increasingly recognized in symptom-level and network approaches to depression measurement. As a retrospective measure administered fortnightly, it is also subject to recall bias and conceals moment-to-moment variability that ecological momentary assessment would better resolve. Fortnightly behavioral aggregation compounds this limitation: screen-by-screen capture at five-second resolution provides far richer temporal information than two-week aggregates can represent, and finer-grained behavioral features may better characterize the timing and patterning of behavioral changes preceding worsening. Richer outcome measures and behavioral features administered at a finer temporal resolution would likely improve both label quality and predictive performance.

With only 37 worsening events in the held-out test set, per-cell subgroup counts are too small for formal inference and performance heterogeneity across transition types is directional rather than confirmatory. Larger samples with higher worsening event counts are needed to truly characterize for whom the approach works better or worse. A descriptive subgroup performance analysis by sex, race, ethnicity, age, and income is reported in Supplementary Table S9; discriminative performance was consistent across subgroups examined (AUC 0.86 - 0.95), though cell sizes preclude formal between-group inference. At approximately 20 fortnightly observations per person, the combination of limited individual time series length and the temporal holdout requirement also constrains person-specific modeling within a forecasting framework, as discussed earlier. Sequence models that learn directly from ordered behavioral and symptom observations, rather than fortnightly aggregates, represent a promising architectural direction as longitudinal datasets reach the scale required for stable training[25]. Approaches that adapt population-level parameters to individual histories with limited observations offer another direction for addressing the person-specific calibration challenge identified here.

## *Conclusions*

Predicting whether a specific individual's depressive symptoms will worsen, remain stable, or improve over the coming weeks is a clinically meaningful problem that remains largely unaddressed in passive sensing research. Using year-long Screenome behavioral data and fortnightly CES-D assessments from 96 adults, we demonstrate that within-person trajectory direction is indeed predictable (AUC 0.906; worsening sensitivity 0.838). This generalizes to individuals not seen during training, and remains informative even when assessments are not up to date.

Each person's typical symptom level is the critical predictive anchor: without it, the most consequential worsening transitions go largely undetected; with it, the model distinguishes acute escalation from chronic baseline. Behavioral features do not independently add predictive power above this clinical anchor, but they characterize how deterioration unfolds in the weeks before threshold crossings, surfacing signals in social media engagement, device use fragmentation, and overnight activity that would otherwise go unobserved between self-report assessments.

These findings establish a conservative lower bound on what passive monitoring systems can accomplish using a self-report measure administered fortnightly and behavioral features aggregated at the same resolution. Richer instruments, finer temporal resolution, and person-specific model calibration represent a promising direction as longitudinal datasets scale. Within-person trajectory prediction, anchored to each person's own symptom history and behavioral patterns, offers a path toward monitoring systems that are sensitive to how a specific individual is changing rather than where they stand relative to a population, a shift in paradigm that may prove as important as the sensing technology itself.

**Methods**

*Study Design, Procedure and Data Collection*

Data were drawn from the Human Screenome Project[17], a series of longitudinal observational studies collecting continuous passive behavioral data and periodic self-report assessments from adults and adolescents in the United States. Participants for this study sample were recruited via the Qualtrics research panel on a rolling basis between June 2020 and November 2021. Eligibility criteria included residing in the United States, being the sole user of an Android smartphone, and providing informed consent and Health Insurance Portability and Accountability Act (HIPAA) authorization prior to enrollment.

Participants installed a custom Screenomics application on their smartphones, which captured screenshots at approximately five-second intervals whenever the device was active, alongside operating system metadata including foreground application, battery state, and screen activity timestamps[15,16,18]. The application required no active user management and operated transparently with a status bar icon indicating that it was recording data. The open-source Stanford Screenomics application is publicly available on GitHub[18].

Participants completed fortnightly self-report surveys throughout the study period. They were compensated $15 for each completed survey and screenome data submission, with potential earnings of up to $390 for full one-year participation. This study was approved by Stanford University's Institutional Review Board (protocol #56430). All procedures were classified as minimal risk. All data were encrypted during transfer and stored on Stanford University's high-risk personal identifiable information and protected health information–compliant servers. Raw screenshots were accessible only to a small subset of trained researchers under strict institutional protocols; derived features used in analysis were deidentified before leaving secure storage. Research staff were trained as mandated reporters under Stanford University and California guidelines, and protocols were in place to refer participants to appropriate professionals or services if concerning content was identified.

*Depressive Symptoms Assessment*

Depressive symptoms were assessed every two weeks using the Center for Epidemiologic Studies Depression Scale (CES-D) [30], a 20-item self-report measure of depressive symptom frequency over the past week. Items are rated on a four-point scale from 0 (rarely or none of the time) to 3 (most or all of the time), with four positively worded items reverse-scored. Total scores range from 0 to 60, with higher scores indicating greater symptom severity. A score of 16 or above is widely used to identify individuals at elevated risk for clinical depression [35]. The CES-D has demonstrated strong psychometric properties across community and clinical samples and has been validated for longitudinal monitoring designs [30,35,41]. Assessments were administered through Qualtrics every two weeks throughout the study period. The fortnightly cadence was chosen to capture meaningful within-person symptom fluctuation over time while maintaining an acceptable participant burden.

*Participants*

Inclusion required a minimum of 10 completed CES-D surveys every two weeks, distributed across the study period (at least 6 surveys in the training split, 2 in validation, and 2 in the test split), and metadata log missingness below 10% across periods with corresponding CES-D data (periods failing this threshold were excluded from feature construction). Survey periods without metadata coverage were excluded from feature construction.

*Behavioral Features derived from Screenome data*

Behavioral features were derived from two data streams captured by the Screenomics platform: operating system metadata logs continuously recording app foreground events, screenshot timestamps, and device state (e.g., screen on/off); and screenshots captured every five seconds when the screen was on. After excluding survey periods with missing metadata log exceeding 10%, features were computed as daily averages within each observation period, defined as the interval from the end of the preceding CES-D assessment to the start of the current one. Specifically, period totals were divided by the number of days with any recorded screen activity to obtain features that were robust to variation in period length across participants and assessments.

Features were organized across five conceptual domains, each grounded in prior evidence linking the corresponding behavioral dimension to depressive symptoms.

**Dosage** captures overall device use intensity through two features. *Mean daily screens* is the average number of screenshots captured per active day; because the Screenomics platform captures one screenshot every five seconds when the device is active, this is a direct proxy for total active screen time. *Active day ratio* is the proportion of days in the period with any recorded screen activity, capturing consistency of engagement. Changes in habitual engagement level have been associated with depressive symptoms in prior passive sensing research, with both withdrawal from and escalation of device use serving as potential signals of shifting motivational states[8,10].

**Fragmentation** captures the structure of device engagement through three features. *Mean daily unique apps* is the average number of distinct applications used per active day. *Mean daily sessions* is the average number of discrete phone use sessions per active day, capturing how frequently a person picks up and uses their device across the day. *Sessions per screen* is the ratio of total phone use sessions to total screenshots in the period; since screenshots proxy screen time, this captures average session length in terms of screen time, with higher values indicating shorter, more fragmented episodes of device engagement. Prior research has found associations between fragmentation and poorer affective states within individuals over time[19].

**Circadian pattern** captures the timing of device use using two features. *Overnight ratio* is the proportion of daily screenshots captured between midnight and 6 AM. The *overnight screens* feature is the average number of screenshots captured between midnight and 6 AM per active day. Separating overnight proportion from overnight count allows the model to distinguish a person whose device use is predominantly nocturnal from one with high overall volume that extends into overnight hours. Late-night phone use is associated with sleep disruption, which is both a cardinal symptom and a prospective risk factor for depressive episodes[31].

**Social media** captures social platform engagement through two complementary features. *Social ratio* is the proportion of daily screenshots captured while using social applications. *Social screens* is the average number of screenshots captured during social media use per active day. Applications were categorized as social media based on Google Play Store market classifications, including popular social media applications such as TikTok and Instagram, messaging applications such as Telegram, and social networking applications. Social media use has been linked to depressive symptoms in prior work, though the direction and interpretation of the association are heterogeneous across individuals and contexts [32,42].

***Content diversity*** captures the visual variety of screens viewed during the period through two indicators. *Mean daily unique apps*, also included in the fragmentation domain as a measure of app-switching breadth, additionally reflects the variety of content types accessed across the period; it is a single feature that serves both conceptual roles. The second feature is *CLIP Embedding Dispersion,* a scalar dispersion metric derived from CLIP embeddings, which are numerical representations of each screenshot's visual content in a high-dimensional vector space, generated from the screenshots recorded during the study using a pre-trained vision-language model[43]. Only this derived scalar was used as a feature, not the raw embedding vectors or screenshots. Dispersion was operationalized as the mean cosine distance of each screenshot's embedding from the mean embedding vector across all screenshots in the period:

$$D_{mean} = \frac{1}{N} \sum_{i=1}^{N} (1 - \cos(x_i, \mu))$$

where N is the number of screenshots in the period, $x_i$ is the CLIP embedding of screenshot i, and $\mu$ is the mean embedding vector across all screenshots captured in the period. Higher values indicate greater visual variety in content viewed during the period. Prior research has shown that people actively sequence content to balance emotional valence during device use, consistent with mood management and emotion regulation processes[33], suggesting that content diversity may reflect the affective range of content engaged over a period.

In addition to the five behavioral domains, *age* (years) and *gender* (encoded as two binary indicators for male and female, with other as the reference category) were included as demographic covariates, given established gender differences in depressive symptom prevalence and patterns[38], and age-related differences in smartphone use patterns and in the association between device use and mood[39].

**Feature construction and prediction timing.** For each behavioral feature, two values were computed for the current observation period: the absolute level during the period, and the change from the immediately preceding period for that individual. The change values are within-person measures of behavioral shift, computed as each individual's own period-to-period difference rather than deviations from a population mean. Together with the most recent CES-D score and three demographic features, these comprised 21 base features. Prior to modeling, Variance Inflation Factors were computed on the training set and features with VIF exceeding 10 were iteratively removed, resulting in the retention of the period-over-period change in active day ratio rather than its level, and the exclusion of overnight screens, which was collinear with mean daily screens and overnight ratio, its component features. The prior period's values for all 17 behavioral features were additionally carried forward as lag features, providing the model with a second behavioral window spanning the preceding observation period. For each participant's first observation, where no prior period exists, lag features were set to zero. Each person's typical symptom level (the mean CES-D score across their training observations) was included as a fixed feature assigned via participant identifier lookup, with no validation or test CES-D values used in its computation. The full model, therefore, comprised 39 features: 21 base features, 17 behavioral lag features, and each person's typical symptom level. All features were computed from the observation period ending immediately before each CES-D assessment, so that no information from the upcoming survey entered feature construction. For tree-based models, count features, rates, and period-over-period deltas were z-score standardized using parameters estimated from training data and applied to validation and test sets without refitting. Bounded ratio features (ranging from 0 to 1) and the most recent CES-D score were left on their original scales, as standardization would not improve their interpretability or comparability. The exception was ElasticNet, for which all features were standardized to ensure comparable regularization. Full feature definitions are provided in Supplementary Table S7.

*Classification Label Design*

The prediction target at each fortnightly assessment was the trajectory direction of each participant's CES-D change over the subsequent two-week interval, defined as a three-class classification: worsening, remaining stable, or improving. Three operationalizations of this outcome were evaluated: crossing absolute CES-D severity thresholds, exceeding each person's own historical variability, and balanced terciles to assess performance under equal class proportions.

*Severity crossing* (sev_crossing; primary) defined trajectory direction based on movement across established CES-D clinical severity thresholds. Scores below 16 were classified as minimal, scores of 16–23 as moderate, and scores of 24 and above as severe[34,35]. An observation was labeled worsening if the CES-D score at the next assessment crossed into a higher severity category, improving if it crossed into a lower category, and stable if it remained within the same category. This operationalization is motivated by clinical utility: a three-point change within the minimal range is clinically inert, whereas crossing from moderate to severe triggers qualitatively different clinical concerns and treatment responses. Severity crossing thus captures directional change at the thresholds that matter for care decisions, rather than statistical fluctuation. Test set label distributions were: improving 11% (N = 44), stable 80% (N = 330), and worsening 9% (N = 37).

*Personalized threshold* (personal_sd) defined trajectory direction relative to each person's within-person variability. For each participant, a personalized threshold was computed from the standard deviation of their CES-D change scores across training observations:

$$threshold_i = k \times SD_i, SD_i = max(SD(\Delta CES\text{-}D_{train,i}), 3.0)$$

where $i$ indexes participants, k = 1.0 sets the stable-zone width at one iSD of CES-D change, with a 3-point floor to prevent near-zero thresholds for low-variability participants. In practice, this means an observation is labeled worsening if the person's CES-D increased by more than one standard deviation above their own typical fluctuation, improving if it decreased by more than one standard deviation, and stable otherwise. This operationalization addresses a limitation of fixed clinical thresholds: for a person whose symptoms are chronically moderate, any given change is evaluated against population-level boundaries that may not reflect what is unusual for that individual. Personalizing the threshold to each person's own historical volatility makes trajectory direction a function of deviation from their own baseline rather than from the population distribution. Person-level SD of CES-D change scores ranged from 3.0 to 22.4 points (population SD = 7.4 points), reflecting substantial heterogeneity in within-person variability. Test set distributions: improving 12% (N = 50), stable 78% (N = 320), worsening 10% (N = 41).

*Balanced tercile* (balanced_tercile) assigned labels by rank-ordering all training-set CES-D change scores and dividing them into three equal-sized bins, with thresholds derived from the training distribution and applied to validation and test sets. Worsening was defined as the top third of CES-D change scores, improving as the bottom third, and stable as the middle third, with ties at bin boundaries broken randomly using a fixed seed. This operationalization enforces balanced class proportions, improving 33% (N = 137), stable 33% (N = 137), and worsening 33% (N = 137) in the test set, eliminating class imbalance as a potential confound and enabling comparison of model performance across labels under identical class priors.

Primary results are reported for severity crossing throughout the manuscript. Detailed results for personalized threshold and balanced tercile are reported in the Supplementary Materials.

*Models*

Four model families were evaluated: ElasticNet logistic regression[44], XGBoost[45], LightGBM[46], and Support Vector Machine[47]. These were selected for their suitability to tabular behavioral features at modest sample sizes, and have demonstrated strong performance in comparable digital health prediction tasks[23,24]. All four models used balanced class weighting to address the approximately 78–80% stable class majority under the severity crossing and personalized threshold labels.

A single population-level model was trained across all participants. The feature set captures within-person dynamics through lag features describing each individual's behavioral history at each prediction point, and through each person's typical symptom level, anchoring predictions to their own chronic baseline.

ElasticNet logistic regression combined L1 and L2 regularization via the SAGA solver with a maximum of 2,000 iterations, with features z-score standardized prior to fitting so that the L1 penalty is applied uniformly across features. XGBoost used the multi:softprob objective for three-class probability output with sample weights computed as inverse class frequencies, equivalent to balanced class weighting. LightGBM used class_weight = 'balanced'. SVM used class_weight = 'balanced' with Platt scaling for calibrated class probabilities; both RBF and linear kernels were evaluated during hyperparameter search. Hyperparameter grids and final selected values are reported in Supplementary Materials and can be found in [our Github repository](#)[48].

*Evaluation Design*

**Data splits.** Observations were divided into training (60%), validation (20%), and test (20%) splits using a temporal within-person design: for each participant, earlier observations were assigned to training, intermediate observations to validation, and later observations to test, strictly preserving chronological order. All 96 participants appear in all three splits. The validation set was used exclusively for hyperparameter selection; all reported performance metrics are from the held-out test set (total of 411 observations). Person-level statistics, including each person's typical symptom level, were computed from training observations only and assigned to validation and test via participant identifier lookup, with no information from validation or test periods used in their computation.

**Primary evaluation: temporal generalization.** The primary evaluation tests whether the model generalizes to future time points for known individuals. Because training periods strictly precede test periods within each person, this design reflects the deployment scenario of a monitoring system predicting upcoming symptom change for an individual with an established assessment history.

**Secondary evaluation: person-level generalization.** A repeated leave-group-out cross-validation was conducted to estimate performance for individuals entirely unseen during training, reflecting the deployment scenario of applying the model to a new person. In each fold, approximately 20% of participants (N ≈ 19) were held out entirely, their data appearing in neither training nor validation. Each person's typical symptom level was set to the population mean for held-out individuals, as no training history was available. This procedure was repeated five times with different participant-to-fold assignments (5 repeats × 5 folds = 25 total evaluations), with fold assignments reshuffled each repeat to ensure every participant was held out multiple times across different training configurations. Performance is summarized as the mean and standard deviation of AUC across the 25 evaluations.

**Metrics.** Four metrics are reported for each model, with 95% confidence intervals derived from 1,000 bootstrap resamples of the test set (seed = 42): area under the receiver operating characteristic curve using a one-vs-rest macro-average (AUC), balanced accuracy, sensitivity for the worsening class, and positive predictive value (PPV) for the worsening class. AUC is the primary metric because it evaluates discriminative ability across all decision thresholds and is robust to class imbalance. Balanced accuracy is the average sensitivity across all three trajectory classes and is reported as the grid-search criterion for comparability with model selection. Worsening-class sensitivity is the proportion of true worsening episodes that the model correctly identified. PPV is the proportion of worsening predictions that were correct, characterizing the false alarm burden more directly than specificity, which is uninformative given the 80% stable majority. The tradeoff between sensitivity and PPV is tunable via the prediction threshold and is discussed in the context of deployment in the Results.

**Feature ablation and significance testing.** To decompose the contribution of each data source, all models were evaluated across four nested feature conditions: (1) the most recently available CES-D score as the sole predictor feature, (2) adding current-period behavioral features, (3) adding behavioral lag features from the prior period, and (4) adding each person's typical symptom level. The significance of each additive step was assessed using a paired percentile bootstrap with 1,000 resamples (seed = 42), testing whether the difference in AUC between adjacent feature conditions exceeded zero. Multiple comparisons across models and label operationalizations were corrected using the Holm-Bonferroni procedure.

**Statistical baselines.** Four rule-based baselines were evaluated to anchor model performance (Tables 2 and 3). The predict-all-stable baseline predicts stable for every observation and represents the performance floor that any useful model must exceed; it is equivalent to the population modal baseline under severity crossing and personalized threshold operationalizations, where stable is the majority class. The person-specific modal baseline predicts whichever trajectory class each person showed most frequently during their training observations. The last-value-carried-forward baseline repeats the previous period's class label. The regression-to-person-mean baseline predicts the direction in which each person's current CES-D severity band would need to move to return to their typical severity band (AUC = 0.750). This last baseline is the strongest non-model comparator and serves as the primary benchmark against which the full model's incremental value is assessed, as it encodes the same within-person symptom anchor without behavioral features or learned parameters.

# References

1. Herrman, H. *et al.* Time for united action on depression: a Lancet-World Psychiatric Association Commission. *Lancet* **399**, 957–1022 (2022).

2. van Eeden, W. A., van Hemert, A. M., Carlier, I. V. E., Penninx, B. W. & Giltay, E. J. Severity, course trajectory, and within-person variability of individual symptoms in patients with major depressive disorder. *Acta Psychiatr. Scand.* **139**, 194–205 (2019).

3. Fried, E. I., Flake, J. & Robinaugh, D. J. Revisiting the theoretical and methodological foundations of depression measurement. *Nat. Rev. Psychol.* **1**, 358–368 (2022).

4. Molenaar, P. C. M. A manifesto on psychology as idiographic science: Bringing the person back into scientific psychology, this time forever. *Measurement (Mahwah NJ)* **2**, 201–218 (2004).

5. Hamaker, E. Why researchers should think 'within-person': A paradigmatic rationale. *Handbook of research methods for studying daily life.* **676**, 43–61 (2012).

6. Torous, J. *et al.* The growing field of digital psychiatry: current evidence and the future of apps, social media, chatbots, and virtual reality. *World Psychiatry* **20**, 318–335 (2021).

7. Torous, J., Kiang, M. V., Lorme, J. & Onnela, J.-P. New tools for new research in psychiatry: A scalable and customizable platform to empower data driven smartphone research. *JMIR Ment. Health* **3**, e16 (2016).

8. Mohr, D. C., Zhang, M. & Schueller, S. M. Personal sensing: Understanding mental health using ubiquitous sensors and machine learning. *Annu. Rev. Clin. Psychol.* **13**, 23–47 (2017).

9. Gillan, C. M. & Rutledge, R. B. Smartphones and the neuroscience of mental health. *Annu. Rev. Neurosci.* **44**, 129–151 (2021).

10. Saeb, S. *et al.* Mobile phone sensor correlates of depressive symptom severity in daily-life behavior: An exploratory study. *J. Med. Internet Res.* **17**, e175 (2015).

11. Jacobson, N. C. & Chung, Y. J. Passive sensing of prediction of moment-to-moment depressed mood among undergraduates with clinical levels of depression sample using smartphones. *Sensors (Basel)* **20**, 3572 (2020).

12. Balliu, B. *et al.* Personalized mood prediction from patterns of behavior collected with smartphones. *NPJ Digit. Med.* **7**, 49 (2024).

13. Webb, C. A. *et al.* Personalized prediction of negative affect in individuals with serious mental illness followed using long-term multimodal mobile phenotyping. *Transl. Psychiatry* **15**, 174 (2025).

14. Vander Zwalmen, Y. *et al.* Mobile technology for just-in-time prediction of depression: a scoping review. *Nat. Ment. Health* (2026) doi:10.1038/s44220-026-00624-6.

15. Reeves, B. *et al.* Screenomics: A Framework to Capture and Analyze Personal Life Experiences and the Ways that Technology Shapes Them. *Hum Comput Interact* **36**, 150–201 (2021).

16. Ram, N. *et al.* Screenomics: A New Approach for Observing and Studying Individuals' Digital Lives. *J. Adolesc. Res.* **35**, 16–50 (2020).

17. Reeves, B., Robinson, T. & Ram, N. Time for the Human Screenome Project. *Nature* **577**, 314–317 (2020).

18. Kim, I. *et al.* An open-source platform for multimodal digital trace data collection from smartphones. *Nat. Health* 1–12 (2026) doi:10.1038/s44360-026-00072-7.

19. Cerit, M. *et al.* Person-specific analyses of smartphone use and mental health: Intensive longitudinal study. *JMIR Form. Res.* **9**, e59875 (2025).

20. Ren, B. *et al.* Predicting states of elevated negative affect in adolescents from smartphone sensors: a novel personalized machine learning approach. *Psychol. Med.* **53**, 5146–5154 (2023).

21. Fisher, H., Nepal, S. & Webb, C. A. Personalized early detection of depression onset using multivariate mobile passive sensing. *Research Square* (2026) doi:10.21203/rs.3.rs-8960944/v1.

22. Amin, R. *et al.* Use of mobile sensing data for longitudinal monitoring and prediction of depression severity: Systematic review. *J. Med. Internet Res.* **27**, e57418 (2025).

23. Leaning, I. E. *et al.* From smartphone data to clinically relevant predictions: A systematic review of digital phenotyping methods in depression. *Neurosci. Biobehav. Rev.* **158**, 105541 (2024).

24. De Angel, V. *et al.* Digital health tools for the passive monitoring of depression: a systematic review of methods. *NPJ Digit. Med.* **5**, 3 (2022).

25. Stamatis, C. A. *et al.* Differential temporal utility of passively sensed smartphone features for depression and anxiety symptom prediction: a longitudinal cohort study. *Npj Ment. Health Res.* **3**, 1 (2024).

26. Zierer, C., Behrendt, C. & Lepach-Engelhardt, A. C. Digital biomarkers in depression: A systematic review and call for standardization and harmonization of feature engineering. *J Affect Disord* **356**, 438–449 (2024).

27. Orben, A. & Przybylski, A. K. The association between adolescent well-being and digital technology use. *Nat. Hum. Behav.* **3**, 173–182 (2019).

28. Ringwald, W. R., King, G., Vize, C. E. & Wright, A. G. C. Passive smartphone sensors for detecting psychopathology. *JAMA Netw. Open* **8**, e2519047 (2025).

29. Orben, A., Meier, A., Dalgleish, T. & Blakemore, S.-J. Mechanisms linking social media use to adolescent mental health vulnerability. *Nat. Rev. Psychol.* **3**, 407–423 (2024).

30. Radloff, L. S. The CES-D scale: a self-report depression scale for research in the general population. *Appl. Psychol. Meas.* **1**, 385–401 (1977).

31. Rod, N. H., Dissing, A. S., Clark, A., Gerds, T. A. & Lund, R. Overnight smartphone use: A new public health challenge? A novel study design based on high-resolution smartphone data. *PLoS One* **13**, e0204811 (2018).

32. Orben, A. Teenagers, screens and social media: a narrative review of reviews and key studies. *Soc. Psychiatry Psychiatr. Epidemiol.* **55**, 407–414 (2020).

33. Cho, M.-J., Reeves, B., Ram, N. & Robinson, T. N. Balancing media selections over time: Emotional valence, informational content, and time intervals of use. *Heliyon* **9**, e22816 (2023).

34. Rushton, J. L., Forcier, M. & Schectman, R. M. Epidemiology of depressive symptoms in the National Longitudinal Study of Adolescent Health. *J. Am. Acad. Child Adolesc. Psychiatry* **41**, 199–205 (2002).

35. Park, S.-H. & Yu, H. Y. How useful is the center for epidemiologic studies depression scale in screening for depression in adults? An updated systematic review and meta-analysis☆. *Psychiatry Res.* **302**, 114037 (2021).

36. Ram, N., Haber, N., Robinson, T. N. & Reeves, B. Binding the Person-Specific Approach to Modern AI in the Human Screenome Project: Moving past Generalizability to Transferability. *Multivariate Behav Res* **59**, 1211–1219 (2024).

37. Kathan, A. *et al.* Personalised depression forecasting using mobile sensor data and ecological momentary assessment. *Front. Digit. Health* **4**, 964582 (2022).

38. Salk, R. H., Hyde, J. S. & Abramson, L. Y. Gender differences in depression in representative national samples: Meta-analyses of diagnoses and symptoms. *Psychol Bull* **143**, 783–822 (2017).

39. Winbush, A. *et al.* Smartphone use in a large US adult population: Temporal associations between objective measures of usage and mental well-being. *Proc Natl Acad Sci U S A* **122**, e2427311122 (2025).

40. Cerit, M. *et al.* Media content atlas: A pipeline to explore and investigate multidimensional media space using multimodal LLMs. in *Proceedings of the Extended Abstracts of the CHI Conference on Human Factors in Computing Systems* 1–13 (ACM, New York, NY, USA, 2025). doi:10.1145/3706599.3720055.

41. Chin, W. Y., Choi, E. P. H., Chan, K. T. Y. & Wong, C. K. H. The psychometric properties of the center for Epidemiologic Studies Depression Scale in Chinese primary care patients: Factor structure, construct

42. validity, reliability, sensitivity and responsiveness. *PLoS One* **10**, e0135131 (2015).

42. Orben, A., Dienlin, T. & Przybylski, A. K. Social media's enduring effect on adolescent life satisfaction. *Proc. Natl. Acad. Sci. U. S. A.* **116**, 10226–10228 (2019).

43. Radford, A. *et al.* Learning transferable visual models from natural language supervision. *ICML* **139**, 8748–8763 (2021).

44. Zou, H. & Hastie, T. Regularization and variable selection via the elastic net. *J. R. Stat. Soc. Series B Stat. Methodol.* **67**, 301–320 (2005).

45. Chen, T. & Guestrin, C. XGBoost: A Scalable Tree Boosting System. in *Proceedings of the 22nd ACM SIGKDD International Conference on Knowledge Discovery and Data Mining* 785–794 (ACM, New York, NY, USA, 2016). doi:10.1145/2939672.2939785.

46. Ke, G. *et al.* LightGBM: A highly efficient gradient Boosting Decision Tree. https://proceedings.neurips.cc/paper/2017/hash/6449f44a102fde848669bdd9eb6b76fa-Abstract.html (2017).

47. Cortes, C. & Vapnik, V. Support-vector networks. *Mach. Learn.* **20**, 273–297 (1995).

48. GitHub - mediacontentatlas/within-person-cesd-screenome. *GitHub*. https://github.com/mediacontentatlas/within-person-cesd-screenome.

# Supplementary Materials

**Table S1** | Classification performance across all four models and all three label operationalizations (held-out test set, N = 411). All three labels are three-class (improving, stable, worsening); AUC is one-vs-rest, macro-averaged. Values are point estimates with 95% bootstrap confidence intervals (1,000 resamples, percentile method, seed = 42) shown below each estimate. Green shading indicates the model with the highest AUC within each label operationalization. All models use the full 39-feature set: base behavioral features, behavioral lag features from the prior period, the most recently available CES-D score, and each person's typical symptom level. Main text Tables 2 and 3 report primary results for severity crossing and personalized threshold; this table provides the complete cross-label picture including balanced tercile. AUC, area under the receiver operating characteristic curve, one-vs-rest macro-average. Balanced accuracy, average sensitivity across all three trajectory classes (worsening, stable, improving). Sensitivity (worsening), proportion of true worsening episodes correctly identified. PPV (worsening), proportion of worsening predictions that were correct. OvR, one-vs-rest; PPV, positive predictive value; CES-D, Center for Epidemiologic Studies Depression Scale.

| Label | Model | AUC (OvR) | Balanced accuracy | Sensitivity, worsening | PPV, worsening |
|---|---|---|---|---|---|
| **Severity crossing** | XGBoost | 0.906 (0.881, 0.929) | 0.834 (0.784, 0.882) | 0.838 (0.707, 0.947) | 0.356 (0.253, 0.464) |
| | LightGBM | 0.901 (0.876, 0.925) | 0.842 (0.798, 0.890) | 0.865 (0.744, 0.970) | 0.344 (0.247, 0.439) |
| | SVM | 0.841 (0.806, 0.876) | 0.696 (0.628, 0.766) | 0.649 (0.488, 0.806) | 0.304 (0.203, 0.405) |
| | ElasticNet | 0.834 (0.797, 0.871) | 0.699 (0.626, 0.764) | 0.730 (0.576, 0.871) | 0.245 (0.167, 0.336) |
| **Personalized threshold** | XGBoost | 0.750 (0.707, 0.793) | 0.620 (0.551, 0.682) | 0.634 (0.485, 0.788) | 0.166 (0.105, 0.224) |
| | LightGBM | 0.731 (0.689, 0.780) | 0.586 (0.524, 0.654) | 0.463 (0.341, 0.585) | 0.145 (0.091, 0.204) |
| | SVM | 0.751 (0.702, 0.796) | 0.620 (0.550, 0.686) | 0.610 (0.454, 0.771) | 0.191 (0.126, 0.257) |
| | ElasticNet | 0.755 (0.704, 0.802) | 0.603 (0.534, 0.669) | 0.561 (0.400, 0.724) | 0.175 (0.110, 0.236) |
| **Balanced tercile** | XGBoost | 0.723 (0.682, 0.764) | 0.555 (0.510, 0.602) | 0.562 (0.479, 0.645) | 0.490 (0.416, 0.566) |
| | LightGBM | 0.732 (0.690, 0.773) | 0.557 (0.511, 0.608) | 0.555 (0.471, 0.642) | 0.481 (0.405, 0.558) |
| | SVM | 0.721 (0.682, 0.760) | 0.557 (0.511, 0.605) | 0.496 (0.411, 0.577) | 0.519 (0.432, 0.606) |
| | ElasticNet | 0.713 (0.672, 0.752) | 0.528 (0.479, 0.575) | 0.474 (0.387, 0.555) | 0.464 (0.388, 0.542) |

**Table S2** | Confusion matrices for all four models under the severity crossing label (held-out test set, N = 411). Rows indicate true class; columns indicate predicted class. Green shading indicates correct classifications (diagonal cells). Row totals reflect true class counts: improving N = 44, stable N = 330, worsening N = 37.

**XGBoost**

| True \ Predicted | Improving | Stable | Worsening | Row total |
|---|---|---|---|---|
| Improving | **40** | 1 | 3 | 44 |
| Stable | 28 | **249** | 53 | 330 |
| Worsening | 1 | 5 | **31** | 37 |
| Col total | 69 | 255 | 87 | 411 |

**LightGBM**

| True \ Predicted | Improving | Stable | Worsening | Row total |
|---|---|---|---|---|
| Improving | **41** | 1 | 2 | 44 |
| Stable | 30 | **241** | 59 | 330 |
| Worsening | 1 | 4 | **32** | 37 |
| Col total | 72 | 246 | 93 | 411 |

**SVM**

| True \ Predicted | Improving | Stable | Worsening | Row total |
|---|---|---|---|---|
| Improving | **33** | 5 | 6 | 44 |
| Stable | 54 | **227** | 49 | 330 |
| Worsening | 5 | 8 | **24** | 37 |
| Col total | 92 | 240 | 79 | 411 |

**ElasticNet**

| True \ Predicted | Improving | Stable | Worsening | Row total |
|---|---|---|---|---|
| Improving | **32** | 3 | 9 | 44 |
| Stable | 45 | **211** | 74 | 330 |
| Worsening | 3 | 7 | **27** | 37 |
| Col total | 80 | 221 | 110 | 411 |

**Table S3** | Moderate-to-severe worsening subgroup analysis (severity crossing label, XGBoost, held-out test set, N = 411). Of the 37 worsening episodes in the test set, 8 involved transitions from moderate (CES-D 16–23) to severe (CES-D 24+). Detection rate at each feature ablation step is shown for three worsening subgroups and the overall worsening class. Each condition adds cumulatively to the prior. Detection is defined as the model assigning predicted class = worsening under the default argmax decision rule. Proportion detected (sensitivity within subgroup) shown in parentheses. Green shading highlights the moderate-to-severe subgroup, which is the focus of the main text claim. The main text statement that the model detects 0% of moderate-to-severe transitions without each person's typical symptom level refers to the 38-feature condition (column 4), the ablation step immediately preceding its addition. CES-D, Center for Epidemiologic Studies Depression Scale.

| Worsening subgroup | N | Prior CES-D only (1 feature) | + Behavioral features (21 features) | + Behavioral lag (38 features) | + Each person's typical symptom level (39 features) |
|---|---|---|---|---|---|
| Minimal → moderate (CES-D <16 → 16–23) | 20 | 15/20 (0.75) | 16/20 (0.80) | 15/20 (0.75) | 15/20 (0.75) |
| Minimal → severe (CES-D <16 → 24+) | 9 | 9/9 (1.00) | 8/9 (0.89) | 8/9 (0.89) | 9/9 (1.00) |
| **Moderate → severe (CES-D 16–23 → 24+)** | 8 | 3/8 (0.38) | 0/8 (0.00) | 0/8 (0.00) | 7/8 (0.88) |
| All worsening (overall) | 37 | 27/37 (0.73) | 24/37 (0.65) | 23/37 (0.62) | 31/37 (0.84) |

**Table S4** | Nested feature ablation for the best-performing model per label operationalization (held-out test set, N = 411). Four cumulative feature conditions are shown: (1) most recently available CES-D score only; (2) adding behavioral features; (3) adding behavioral lag features; (4) adding each person's typical symptom level. AUC is one-vs-rest, macro-averaged, with 95% bootstrap confidence intervals (1,000 resamples, seed = 42). ΔAUC shows the paired-bootstrap mean gain over the previous condition with its 95% CI; p-values are the proportion of bootstrap resamples in which the gain was ≤ 0. Holm-Bonferroni correction is applied over the three-test family comprising the step-(4) transition for each label; ** indicates Holm-corrected p < 0.05, matching the criterion in Figure 2. Steps (2) and (3) report raw p-values for transparency; they are not part of the corrected family. The ElasticNet personalized threshold step-(4) CI lower bound sits at the boundary of zero (δ = +0.052, [-0.001, +0.102]); the Holm-corrected p = 0.030 nonetheless supports the conclusion that the gain is reliable across operationalizations. Models shown are identical to those in Figure 2. AUC, area under the receiver operating characteristic curve; OvR, one-vs-rest; CES-D, Center for Epidemiologic Studies Depression Scale.

| Feature condition | AUC [95% CI] | ΔAUC vs. previous [95% CI] | p (Holm) | Sig. |
|---|---|---|---|---|
| **Severity crossing — XGBoost** | | | | |
| (1) Most recent CES-D only | 0.872 [0.840, 0.903] | — | — | |
| (2) + Behavioral features | 0.876 [0.846, 0.905] | +0.004 [−0.010, +0.017] | 0.298 | |
| (3) + Behavioral lag features | 0.875 [0.844, 0.904] | −0.002 [−0.009, +0.006] | 0.652 | |
| **(4) + Each person's typical symptom level** | 0.906 [0.881, 0.929] | +0.031 [+0.008, +0.057] | 0.018 | ** |
| **Personalized threshold — ElasticNet** | | | | |
| (1) Most recent CES-D only | 0.695 [0.662, 0.727] | — | — | |
| (2) + Behavioral features | 0.707 [0.671, 0.752] | +0.013 [−0.022, +0.045] | 0.228 | |
| (3) + Behavioral lag features | 0.704 [0.663, 0.745] | −0.003 [−0.024, +0.020] | 0.633 | |
| (4) + Each person's typical symptom level | **0.755** [0.704, 0.802] | **+0.052** [−0.001, +0.102] | **0.030** | ** |
| **Balanced tercile — LightGBM** | | | | |
| (1) Most recent CES-D only | 0.642 [0.605, 0.679] | — | — | |
| (2) + Behavioral features | 0.658 [0.620, 0.699] | +0.016 [−0.019, +0.051] | 0.188 | |
| (3) + Behavioral lag features | 0.640 [0.599, 0.680] | −0.018 [−0.044, +0.006] | 0.933 | |
| (4) + Each person's typical symptom level | **0.732** [0.690, 0.773] | **+0.091** [+0.064, +0.120] | **0.000** | ** |

**Table S5** | Two-feature model vs. full 39-feature model (XGBoost, severity crossing label, held-out test set, N = 411). The two-feature model includes only the most recently available CES-D score and each person's typical symptom level; hyperparameters were grid-searched on the validation set using the same search space as the full model. AUC is one-vs-rest, macro-averaged. Values are point estimates with 95% bootstrap confidence intervals (1,000 resamples, seed = 42) shown in parentheses. Green shading in the performance table indicates the metric with the largest difference between models. Prediction agreement is the proportion of test cases on which both models produce identical predicted class. The cross-tabulation diagonal (green) shows matching predictions between models. AUC, area under the receiver operating characteristic curve; OvR, one-vs-rest; PPV, positive predictive value; CES-D, Center for Epidemiologic Studies Depression Scale.

**Performance comparison**

| Metric | 2-feature model | 39-feature model (full) |
|---|---|---|
| AUC (OvR) | **0.917** (0.894, 0.939) | **0.906** (0.881, 0.929) |
| Balanced accuracy | 0.825 (0.773, 0.875) | 0.834 (0.784, 0.882) |
| Sensitivity, worsening | 0.811 (0.682, 0.935) | 0.838 (0.707, 0.947) |
| PPV, worsening | 0.353 (0.253, 0.458) | 0.356 (0.253, 0.464) |
| Prediction agreement | 0.990 (407/411) (0.981, 0.998) | — |

**Prediction agreement by true class**

| True class | N | Prediction agreement |
|---|---|---|
| Improving | 44 | 44/44 (1.000) |
| Stable | 330 | 329/330 (0.997) |
| Worsening | 37 | 34/37 (0.919) |

**Prediction cross-tabulation (2-feature vs. 39-feature)**

| 2-feat  39-feat | Improving | Stable | Worsening | Row total |
|---|---|---|---|---|
| Improving | 68 | 0 | 3 | 71 |
| Stable | 0 | 255 | 0 | 255 |
| Worsening | 1 | 0 | 84 | 85 |

**Table S6** | Fold-by-fold results across the 5 × 5 leave-group-out cross-validation procedure (25 total evaluations; XGBoost, severity crossing label, 39-feature model). In each fold, approximately 19–20 participants were held out entirely from training; each person's typical symptom level was set to the population mean of training participants for held-out individuals. The full model was refit from scratch for each fold using deterministic seeds. AUC is one-vs-rest, macro-averaged. Yellow shading indicates the one fold in which no worsening predictions were made (Repeat 5, Fold 4; 3 true worsening episodes in that held-out group). Summary values reported in main text Table 4 are the mean across all 25 fold-level evaluations; the interval in Table 4 is the minimum and maximum of the five per-repeat means shown in the summary table below. AUC, area under the receiver operating characteristic curve; OvR, one-vs-rest; PPV, positive predictive value.

**Fold-by-fold results**

| Repeat | Fold | Held-out participants | Held-out observations | Train (n) | Val (n) | AUC | Balanced accuracy | Sensitivity, worsening | PPV, worsening |
|---|---|---|---|---|---|---|---|---|---|
| 1 | 1 | 20 | 89 | 936 | 310 | 0.820 | 0.740 | 0.800 | 0.222 |
| 1 | 2 | 19 | 81 | 967 | 320 | 0.834 | 0.685 | 0.333 | 0.400 |
| 1 | 3 | 19 | 75 | 974 | 322 | 0.800 | 0.750 | 0.667 | 0.111 |
| 1 | 4 | 19 | 84 | 953 | 314 | 0.823 | 0.686 | 0.462 | 0.353 |
| 1 | 5 | 19 | 82 | 954 | 314 | 0.803 | 0.691 | 0.600 | 0.083 |
| 2 | 1 | 20 | 86 | 942 | 310 | 0.822 | 0.733 | 0.667 | 0.148 |
| 2 | 2 | 19 | 84 | 954 | 316 | 0.863 | 0.799 | 0.800 | 0.190 |
| 2 | 3 | 19 | 82 | 963 | 318 | 0.896 | 0.824 | 0.833 | 0.185 |
| 2 | 4 | 19 | 79 | 968 | 320 | 0.779 | 0.650 | 0.545 | 0.194 |
| 2 | 5 | 19 | 80 | 957 | 316 | 0.802 | 0.637 | 0.333 | 0.429 |
| 3 | 1 | 20 | 85 | 943 | 308 | 0.939 | 0.932 | 1.000 | 0.333 |
| 3 | 2 | 19 | 83 | 955 | 319 | 0.760 | 0.686 | 0.667 | 0.133 |
| 3 | 3 | 19 | 81 | 963 | 319 | 0.823 | 0.806 | 0.800 | 0.400 |
| 3 | 4 | 19 | 79 | 964 | 319 | 0.749 | 0.618 | 0.125 | 0.071 |
| 3 | 5 | 19 | 83 | 959 | 315 | 0.822 | 0.677 | 0.556 | 0.238 |
| 4 | 1 | 20 | 89 | 940 | 310 | 0.909 | 0.867 | 1.000 | 0.308 |
| 4 | 2 | 19 | 80 | 963 | 320 | 0.816 | 0.729 | 0.600 | 0.150 |
| 4 | 3 | 19 | 80 | 961 | 317 | 0.868 | 0.767 | 0.571 | 0.500 |
| 4 | 4 | 19 | 81 | 955 | 315 | 0.737 | 0.572 | 0.182 | 0.125 |
| 4 | 5 | 19 | 81 | 965 | 318 | 0.821 | 0.674 | 0.500 | 0.107 |
| 5 | 1 | 20 | 82 | 949 | 311 | 0.819 | 0.777 | 0.800 | 0.286 |
| 5 | 2 | 19 | 86 | 952 | 316 | 0.835 | 0.758 | 0.700 | 0.389 |
| 5 | 3 | 19 | 83 | 956 | 316 | 0.839 | 0.672 | 0.250 | 0.100 |
| 5 | 4 | 19 | 83 | 956 | 315 | 0.805 | 0.599 | 0.000 | 0.000 |
| 5 | 5 | 19 | 77 | 971 | 322 | 0.743 | 0.661 | 0.429 | 0.136 |

**Per-repeat summary and overall**

| Repeat | AUC | Balanced accuracy | Sensitivity, worsening | PPV, worsening |
|---|---|---|---|---|
| 1 | 0.816 | 0.710 | 0.572 | 0.234 |
| 2 | 0.832 | 0.729 | 0.636 | 0.229 |
| 3 | 0.819 | 0.744 | 0.630 | 0.235 |
| 4 | 0.830 | 0.722 | 0.571 | 0.238 |
| 5 | 0.808 | 0.693 | 0.436 | 0.182 |
| Mean (reported in Table 4) | 0.821 | 0.720 | 0.569 | 0.224 |
| Per-repeat range (reported in Table 4) | (0.808, 0.832) | (0.693, 0.744) | (0.436, 0.636) | (0.182, 0.238) |

**Table S7** | Complete feature definitions for all 39 model inputs, organized by domain. The 39 features comprise 21 base features (1 CES-D, 3 demographic, 17 behavioral), 17 behavioral lag features (prior-period values of all behavioral base features), and each person's typical symptom level. Two candidate features were excluded prior to modeling via VIF screening (VIF > 10): the level of active day ratio and overnight screens. Full feature construction details are in the Methods; this table provides operational definitions and preprocessing notes for reproducibility. Δ denotes period-over-period change from the preceding fortnightly window. CLIP, Contrastive Language-Image Pretraining; CES-D, Center for Epidemiologic Studies Depression Scale; VIF, Variance Inflation Factor.

| Feature | Domain | Computation | Notes |
|---|---|---|---|
| **CES-D features** | | | |
| Most recently available CES-D score | CES-D | CES-D total score from the assessment immediately preceding the current observation period (prior_cesd). Range 0–60. | |
| Each person's typical symptom level | CES-D | Mean CES-D score across all training-set observations for that participant (person_mean_cesd), assigned via participant ID lookup. | Computed from training data only; no leakage from validation or test periods. Cold-start participants receive the population training mean. Fixed across all periods for a given participant. |
| **Demographic features** | | | |
| Age | Demographic | Participant age in years at enrollment. | Continuous. |

| Gender (two binary indicators) | Demographic | Two one-hot indicators encoding self-reported gender: gender_mode_1 (male), gender_mode_2 (female). Reference category: other. | The third category (other; n = 2 participants) dropped. |

**Behavioral features — Dosage**

| Mean daily screens | Dosage | Total screenshots in the period divided by active days (mean_daily_screens). One screenshot ≈ 5 seconds of screen-on time. | Direct proxy for total active screen time. Level and Δ included. Lag versions included. |
|---|---|---|---|
| Active day ratio (Δ only) | Dosage | Period-over-period change in the proportion of days with any recorded screen activity (active_day_ratio_delta). | Level excluded after VIF screening (collinear with dosage features). Only delta retained. Lag version included. |

**Behavioral features — Fragmentation**

| Mean daily unique apps | Fragmentation / Content diversity | Mean distinct foreground applications per active day (mean_daily_unique_apps). | Single feature serving two conceptual roles: app-switching breadth and content variety. Level and Δ included. Lag versions included. |
|---|---|---|---|
| Mean daily sessions | Fragmentation | Mean discrete phone use sessions per active day (mean_daily_switches). | Higher values indicate more frequent pick-up-and-use episodes. Level and Δ included. Lag versions included. |
| Sessions per screen | Fragmentation | Ratio of total phone use sessions to total screenshots in the period (switches_per_screen). | Higher values indicate shorter, more fragmented episodes. Level and Δ included. Lag versions included. |

**Behavioral features — Circadian pattern**

| Overnight ratio | Circadian pattern | Proportion of daily screenshots captured between midnight and 6 AM (mean_daily_overnight_ratio). | Bounded [0, 1]. Level and Δ included. Lag versions included. Overnight screens excluded after VIF screening (collinear with mean daily screens and overnight ratio). |
|---|---|---|---|

**Behavioral features — Social media**

| Social ratio | Social media | Proportion of daily screenshots during social applications (mean_daily_social_ratio). Apps classified using Google Play Store market categories. | Bounded [0, 1]. Level and Δ included. Lag versions included. |
|---|---|---|---|
| Social screens | Social media | Mean screenshots during social media use per active day (mean_daily_social_screens). Computed as mean_daily_screens × mean_daily_social_ratio. | Level and Δ included. Lag versions included. |

**Behavioral features — Content diversity**

| CLIP embedding dispersion | Content diversity | Mean cosine distance of each screenshot's CLIP embedding from the period mean: $(1/N) \Sigma (1 - \cos(v\_i, \mu))$, where $v\_i$ is the embedding of screenshot i and $\mu$ is the period mean. | Derived from a pre-trained vision-language model (CLIP; Radford et al., 2021). Only the scalar dispersion enters as a feature. Level and Δ included. Lag versions included. |
|---|---|---|---|
| **Feature construction and preprocessing notes** | | | |
| Lag features | All behavioral | Prior-period values of all 17 behavioral base features carried forward as additional inputs. | First-observation lags set to zero. Static demographics, prior_cesd, and cesd_delta excluded from lag set (time-invariant or outcome-adjacent). |
| VIF screening | All features | Variance Inflation Factors computed on training set; features with VIF > 10 iteratively removed. | Two features removed: active_day_ratio (level) and mean_daily_overnight_screens_delta. Their delta and component features were retained. |
| Standardization | All features | For ElasticNet, all the features are Z-score standardized using training-set parameters, frozen for validation and test sets. | |
| Temporal alignment | All features | All current-period features summarize the fortnightly window between CES-D assessments. Lag features summarize the same window one period earlier. | No information from the upcoming CES-D assessment enters feature construction. |

**Table S8** | Final hyperparameter values selected by grid search on the validation set for all four models and all three label operationalizations (full 39-feature model). Grid search used balanced accuracy as the selection criterion. These values produced the test-set performance reported in main text Tables 2 and 3 and Supplementary Tables S1 and S4. All ElasticNet fits use penalty = elasticnet, solver = saga, max_iter = 2000, class_weight = balanced, random_state = 42, with features z-score standardized prior to fitting so that L1 regularization applies a uniform penalty across features. All XGBoost fits use objective = multi:softprob, num_class = 3, class weights derived from inverse training class frequencies, random_state = 42. All LightGBM fits use class_weight = balanced, random_state = 42. All SVM fits use class_weight = balanced, probability = True (Platt scaling), random_state = 42, with features standardized prior to fitting.

| Label | Model | Selected hyperparameters |
|---|---|---|
| **Severity crossing** | ElasticNet | C = 0.1, l1_ratio = 0.9 |
| | XGBoost | learning_rate = 0.01, max_depth = 3, n_estimators = 100<br>min_child_weight = 1, subsample = 1.0, colsample_bytree = 1.0 |
| | LightGBM | learning_rate = 0.01, max_depth = 3, n_estimators = 50, num_leaves = 31<br>min_child_samples = 30, subsample = 0.8, colsample_bytree = 1.0<br>reg_alpha = 0.0, reg_lambda = 0.1 |
| | SVM | kernel = rbf, C = 10.0, gamma = 0.001 |
| | ElasticNet | C = 0.5, l1_ratio = 0.99 |
| | XGBoost | learning_rate = 0.01, max_depth = 5, n_estimators = 100<br>min_child_weight = 3, subsample = 1.0, colsample_bytree = 1.0 |

| | | | |
|---|---|---|---|
| **Personalized threshold** | LightGBM | learning_rate = 0.01, max_depth = 5, n_estimators = 100, num_leaves = 15<br>min_child_samples = 30, subsample = 0.8, colsample_bytree = 1.0<br>reg_alpha = 1.0, reg_lambda = 0.1 | |
| | SVM | kernel = linear, C = 5.0, gamma = 0.0001 | |
| **Balanced tercile** | ElasticNet | C = 0.5, l1_ratio = 0.5 | |
| | XGBoost | learning_rate = 0.05, max_depth = 5, n_estimators = 100<br>min_child_weight = 3, subsample = 1.0, colsample_bytree = 1.0 | |
| | LightGBM | learning_rate = 0.05, max_depth = 5, n_estimators = 50, num_leaves = 15<br>min_child_samples = 10, subsample = 0.8, colsample_bytree = 1.0<br>reg_alpha = 1.0, reg_lambda = 0.1 | |
| | SVM | kernel = linear, C = 0.5, gamma = 0.0001 | |

**Table S9** | Subgroup performance analysis of the full 39-feature XGBoost model under the severity crossing label, disaggregated by participant demographics on the held-out test set (N = 411 observations, 96 participants, 37 worsening events). Point estimates are reported with 95% bootstrap confidence intervals (1,000 observation-level resamples, seed = 42, percentile method) shown in parentheses. Predictions are identical to those used in main text Table 2 and Figure 3; no refitting, subgroup-specific training, or threshold tuning was performed. Subgroups are derived from the demographic survey summarized in main text Table 1. Cell sizes for worsening events are small (8 to 27 per cell), and confidence intervals are correspondingly wide; between-subgroup differences should be interpreted descriptively. No between-subgroup hypothesis tests are reported. AUC, area under the receiver operating characteristic curve, one-vs-rest macro-average. Balanced accuracy, average sensitivity across all three trajectory classes (worsening, stable, improving). Sensitivity (worsening), proportion of true worsening episodes correctly identified. PPV (worsening), proportion of worsening predictions that were correct. OvR, one-vs-rest; PPV, positive predictive value.

| Subgroup axis | Group | N participants | N test obs | N worsening | AUC (OvR) | Balanced accuracy | Sensitivity, worsening | PPV, worsening |
|---|---|---|---|---|---|---|---|---|
| Overall | All participants | 96 | 411 | 37 | 0.906 (0.881, 0.929) | 0.834 (0.784, 0.882) | 0.838 (0.707, 0.947) | 0.356 (0.253, 0.464) |
| Sex | Female | 56 | 243 | 27 | 0.910 (0.874, 0.941) | 0.827 (0.755, 0.891) | 0.815 (0.650, 0.962) | 0.379 (0.241, 0.509) |
| | Male | 38 | 161 | 10 | 0.908 (0.865, 0.945) | 0.875 (0.793, 0.935) | 0.900 (0.692, 1.000) | 0.333 (0.167, 0.520) |
| Race [a] | White | 62 | 267 | 22 | 0.907 (0.876, 0.935) | 0.843 (0.782, 0.897) | 0.909 (0.765, 1.000) | 0.328 (0.219, 0.456) |
| | Non-White | 34 | 144 | 15 | 0.906 (0.864, 0.946) | 0.826 (0.730, 0.906) | 0.733 (0.499, 0.933) | 0.423 (0.231, 0.630) |
| Ethnicity | Non-Hispanic | 80 | 345 | 28 | 0.911 (0.882, 0.937) | 0.830 (0.769, 0.888) | 0.821 (0.667, 0.957) | 0.359 (0.246, 0.492) |
| | Hispanic/Latinx | 16 | 66 | 9 | 0.857 (0.785, 0.916) | 0.834 (0.741, 0.905) | 0.889 (0.625, 1.000) | 0.348 (0.167, 0.552) |
| Age tercile | Younger (20–37) | 32 | 129 | 18 | 0.888 (0.842, 0.929) | 0.806 (0.719, 0.887) | 0.833 (0.647, 1.000) | 0.395 (0.250, 0.548) |
| | Middle (37–57) | 33 | 143 | 8 | 0.877 (0.813, 0.931) | 0.781 (0.644, 0.893) | 0.750 (0.400, 1.000) | 0.214 (0.069, 0.381) |
| | Older (57+) | 31 | 139 | 11 | 0.946 (0.910, 0.974) | 0.916 (0.846, 0.962) | 0.909 (0.700, 1.000) | 0.476 (0.250, 0.696) |
| Household income [b] | < $25,000 | 38 | 161 | 17 | 0.887 (0.844, 0.926) | 0.816 (0.731, 0.888) | 0.882 (0.706, 1.000) | 0.341 (0.200, 0.500) |
| | $25,000–$99,999 | 22 | 98 | 9 | 0.899 (0.838, 0.948) | 0.814 (0.694, 0.919) | 0.778 (0.500, 1.000) | 0.350 (0.167, 0.579) |
| | ≥ $100,000 | 30 | 127 | 8 | 0.931 (0.880, 0.974) | 0.828 (0.697, 0.947) | 0.750 (0.429, 1.000) | 0.333 (0.120, 0.562) |

[a] "Non-White" aggregates participants who identified with any of Black or African American, Asian, Native American or Alaska Native, Pacific Islander, Other, or any multiracial combination. Two participants with self-reported gender "Other" (9 test observations) are excluded from between-group comparison but included in the overall row.

[b] Six participants (24 test observations) who responded "Prefer not to answer or don't know" to household income are excluded from between-group comparison but included in the overall row.

**Data availability statement**

All deidentified and aggregated data used and reported in this study are available from the corresponding author upon reasonable request, subject to institutional data sharing agreements and IRB restrictions.

**Code availability statement**

The Stanford Screenomics data collection platform is open source and available on Github at https://github.com/StanfordScreenomics/Platform.

Analysis code for all models and evaluations reported in this study is open source and available on Github at https://github.com/mediacontentatlas/within-person-cesd-screenome.

**Author contributions (CRediT format)**

T.N.R., B.R., and N.R.: Human Screenome Project conceptualization, data collection supervision. M.C. and N.H.: conceptualization of the current study. M.C.: data curation and cleaning, methodology, analysis, investigation. A.M., and V.A.F. (equal contribution): data curation, cleaning and analysis contributions. M.C.: writing, original draft. T.N.R., B.R., N.R., and N.H.: supervision and review. All authors: writing, review, and editing, final approval.

**Competing interests**

The authors declare no competing interests.

**Acknowledgements**

Research reported in this publication was supported, in part, by the National Heart, Lung, and Blood Institute of the National Institutes of Health under Award Number R01HL169601, the Stanford Maternal and Child Health Research Institute (MCHRI), Stanford University, the Department of Pediatrics, Stanford University, a Stanford Predictives and Diagnostics Accelerator (SPADA) grant from the Stanford Center for Clinical and Translational Research and Education (SPECTRUM), Stanford University, and the Stanford Institute for Human-Centered Artificial Intelligence (HAI), Stanford University. The content is solely the responsibility of the authors and does not necessarily represent the official views of the National Institutes of Health or other funders.